\documentclass[pdf]{statsoc}

\usepackage{geometry} 
\usepackage{vmargin} 
\usepackage{amsmath,amssymb}
\usepackage{graphicx}
\usepackage{pdflscape}
\usepackage{bm}
\usepackage{multirow,bigdelim}
\usepackage{color} 
\usepackage{soul}
\usepackage{url}
\usepackage{textcomp}

\newcommand{\dquote}{\textquotesingle\textquotesingle}

\title[Statistical Modelling of Citation Exchange]{Statistical Modelling of Citation Exchange Between \\ Statistics Journals}

\author{Cristiano Varin}
\address{Universit\`a Ca' Foscari,
         Venezia,
         Italy.}
\email{sammy@unive.it}
\author[C. Varin, M. Cattelan, and D. Firth]{Manuela Cattelan}
\address{Universit\`a degli Studi di Padova,
Padova,
Italy.}
\author[C. Varin, M. Cattelan, and D. Firth]{David Firth}
\address{University of Warwick,
Coventry,
UK.}

\begin{document}

\begin{abstract}
Rankings of scholarly journals based on citation data are often met with skepticism by the scientific community. Part of the skepticism is due to disparity between the common perception of journals' prestige and their ranking based on citation counts. A more serious concern is the inappropriate use of journal rankings to evaluate the scientific influence of authors. This paper focuses on analysis of the table of cross-citations among a selection of Statistics journals. Data are collected from the Web of Science database published by Thomson Reuters. Our results suggest that modelling the exchange of citations between journals is useful to highlight the most prestigious journals, but also that journal citation data are characterized by considerable heterogeneity, which needs to be properly summarized. Inferential conclusions require care in order to avoid potential over-interpretation of insignificant differences between journal ratings.  Comparison with published ratings of institutions from the UK's Research Assessment Exercise shows strong correlation at aggregate level between assessed research quality and journal citation `export scores' within the discipline of Statistics.

\end{abstract}

\section{Introduction}\label{sect:intro}

The problem of ranking scholarly journals has arisen partly as an economic matter. When the number of scientific journals started to increase, librarians were faced with decisions as to which journal subscriptions should consume their limited economic resources; a natural response was to be guided by the relative importance of different journals according to a published or otherwise agreed ranking. \citet{gross_27} proposed the counting of citations received by journals as a direct measure of their importance. \citet{garfield_55} suggested that the number of citations received should be normalized by the number of citable items published by a journal. This idea is at the origin of the \emph{Impact Factor}, the best known index for ranking journals. Published since the 1960s, the Impact Factor is `an average citation rate per published article' \citep{garfield_72}.

The Impact Factor of the journals where scholars publish has also been employed --- improperly, many might argue --- in appointing to academic positions, in awarding research grants, and in ranking universities and their departments.  
The \emph{San Francisco Declaration on Research Assessment} (\citeyear{dora_13}) and the IEEE Position Statement on \emph{Appropriate Use of Bibliometric Indicators for the Assessment of Journals, Research Proposals, and Individuals} \citep{ieee_13} are just two of the most recent authoritative standpoints regarding the risks of automatic, metric-based evaluations of scholars. 
Typically, only a small fraction of all published articles accounts for most of the citations received by a journal \citep{seglen_97}.  Single authors should ideally be evaluated on the basis of their own outputs and not through citations of other papers that have appeared in the journals where their papers have been published \citep{seglen_97, adler_09, silverman_09}. As stated in a recent \emph{Science} editorial about Impact Factor distortions \citep{alberts_13}:
\begin{quote}
`(\ldots) the leaders of the scientific enterprise must accept full responsibility for thoughtfully analyzing the scientific contributions of other researchers. To do so in a meaningful way requires the actual reading of a small selected set of each researcher's publications, a task that must not be passed by default to journal editors'.
\end{quote}
Indicators derived from citations received by papers written by a particular author (\emph{e.g.}, \citeauthor{bornmann_14}, \citeyear{bornmann_14})  can be a useful complement for evaluation of trends and patterns of that author's impact, but not a substitute for the reading of papers. 

Journal rankings based on the Impact Factor often differ substantially from common perceptions of journal prestige \citep{theoharakis_03, arnold_11}. Various causes of such discrepancy have been pointed out. First, there is the phenomenon that more `applied' journals tend to receive citations from other scientific fields more often than do journals that publish theoretical work. This may be related to uncounted `indirect citations' arising when methodology developed in a theoretical journal is then popularized by papers published in applied journals accessible to a wider audience and thus receiving more citations than the original source \citep{journal-ranking_07, putirka_13}. 
Second is the short time-period used for computation of the Impact Factor, which can be completely inappropriate for some fields, in particular for Mathematics and Statistics \citep{nierop_09, arnold_11}\null. Finally, there is the risk of manipulation, whereby authors might be asked by journal editors to add irrelevant citations to other papers published in their journal \citep{sevinc_04, frandsen_07, archambault_09, arnold_11}\null.  According to a large survey published in \emph{Science} \citep{wilhite_12}, about $20\%$ of academics in social-science and business fields have been asked to `pad their papers with superfluous references to get published' \citep{vannorden_12}. The survey data also suggest that junior faculty members are more likely to be pressured to cite superfluous papers. Recently, Thomson Reuters has started publishing the Impact Factor  both {with} and {without} journal self-citations, thereby allowing evaluation of the contribution of self-citations to the Impact Factor calculation.  Moreover, Thomson Reuters has occasionally excluded journals with an excessive self-citation rate from the Journal Citation Reports.  

Given the above criticisms, it is not surprising that the Impact Factor and other `quantitative' journal rankings have given rise to substantial skepticism about the value of citation data. Several proposals have been developed in the bibliometric literature to overcome the weaknesses of the Impact Factor; examples include the \emph{Article Influence Score}  \citep{bergstrom_07, west_10}, the \emph{H index} for journals \citep{braun_06, pratelli_12}, the {\it Source Normalized Impact per Paper} (SNIP) index \citep{waltman_13}, and methods based on percentile rank classes \citep{marx_13}.

The aforementioned \emph{Science} editorial  \citep{alberts_13} reports that 
\begin{quote}
`(...) in some nations, publication in a journal with an impact factor below $5.0$ is officially of zero value.'
\end{quote}
In the latest edition (2013) of the Journal Citation Reports, the only journal with an Impact Factor larger than 5 in the category \emph{Statistics and Probability} was the \emph{Journal of the Royal Statistical Society} Series B, with Impact Factor 5.721.  The category \emph{Mathematics} achieved still lower Impact Factors, with the highest value there in 2013 being $3.08$ for \emph{Communications on Pure and Applied Mathematics}. Several bibliometric indicators have been developed, or adjusted, to allow for cross-field comparisons, \emph{e.g.}~\cite{leydesdorff_13}, \cite{waltman_13b}, and could be considered to alleviate unfair comparisons. However, our opinion is that comparisons between different research fields will rarely make sense, and that such comparisons should be avoided.  Research fields differ very widely, for example in terms of the frequency of publication, the typical number of authors per paper and the typical number of citations made in a paper, as well as in the sizes of their research communities. Journal homogeneity is a minimal prerequisite for a meaningful statistical analysis of citation data \citep{lehmann_09}. 

Journal citation data are unavoidably characterized by substantial variability \citep[\emph{e.g.}, ][]{amin_00}. A clear illustration of this variability, suggested by the Associate Editor of this paper, comes from an early editorial of \emph{Briefings in Bioinformatics} \citep{bishop_07}  announcing that this new journal had received an Impact Factor of $24.37$\null. However, the editors noted that a very large fraction of the journal's citations came from a single paper; if that paper were to be dropped, then the journal's Impact Factor would decrease to about $4$. The variability of the Impact Factor is inherently related to the heavy-tailed distribution of citation counts. Averaged indicators such as the Impact Factor are clearly unsuitable for summarizing highly skew distributions. Nevertheless, quantification of uncertainty is typically lacking in published rankings of journals. A recent exception is \cite{chen_14} who employ a bootstrap estimator for the variability of journal Impact Factors.   Also the SNIP indicator published by Leiden University's Centre for Science and Technology Studies 
based on the Elsevier Scopus database, and available online at \url{www.journalindicators.com},  is accompanied by a `stability interval' computed via a bootstrap method. See also \citeauthor{hall_09a} (\citeyear{hall_09a}, \citeyear{hall_10}) and references therein for more details  on statistical assessment of the authority of rankings. 

The Impact Factor was developed to identify which journals have the greatest influence on subsequent research. The other metrics mentioned in this paper originated as possible improvements on the Impact Factor, with the same aim. \cite{palacios_04} list a set of properties that a ranking method which measures the intellectual influence of journals, by using citation counts, should satisfy. However, the list of all desirable features of a ranking method might reasonably be extended to include features other than citations, depending on the purpose of the ranking.  For example, when librarians decide which journals to take, they should consider the influence of a journal in one or more research fields, but they may also take into account its cost-effectiveness. The website \texttt{www.journalprices.com}, maintained by Professors Ted Bergstrom and Preston McAfee, ranks journals according to their price per article, price per citation, and a composite index. 

A researcher when deciding where to submit a paper most likely considers each candidate journal's record of publishing papers on similar topics, and the importance of the journal in the research field; but he/she may also consider the speed of the reviewing process, the typical time between acceptance and publication of the paper, possible page charges, and the likely effect on his/her own career.  Certain institutions and national evaluation agencies publish rankings of journals which are used to evaluate researcher performance and to inform the hiring of new faculty members. 
For various economics and management-related disciplines, the \emph{Journal Quality List}, compiled by Professor Anne-Wil Harzing and available at \texttt{www.harzing.com/jql.htm}, combines more than 20 different rankings made by universities or evaluation agencies in various countries. Such rankings typically are based on bibliometric indices, expert surveys, or a mix of both.

Modern technologies have fostered the rise of alternative metrics such as ``webometrics'' based on citations on the internet or numbers of downloads of articles. Recently, interest has moved from web-citation analysis to social-media usage analysis. In some disciplines the focus is now towards broader measurement of research impact through the use of web-based quantities such as citations in social-media sites, newspapers, government policy documents, blogs, etc. This is mainly implemented at the level of individual articles, see for example the Altmetric service \citep{adie_13} available at \texttt{www.altmetric.com}, but the analysis may also be made at journal level.  Along with the advantages of timeliness, availability of data and consideration of different sources, such measures also have certain drawbacks related to data quality, possible bias, and data manipulation \citep{bornmann2_14}.

A primary purpose of the present paper is to illustrate the risks of over-interpretation of insignificant differences between journal ratings. In particular, we focus on the analysis of the exchange of citations among a relatively homogeneous list of journals. Following \cite{stigler_94}, we model the table of cross-citations between journals in the same field by using a Bradley-Terry model \citep{bradley_52} and thereby derive a ranking of the journals' ability to `export intellectual influence' \citep{stigler_94}. Although the Stigler approach has desirable properties and is simple enough to be promoted also outside the statistics community,  there have been rather few published examples of application of this model since its first appearance; \cite{stigler_95} and  \cite{liner_04} are two notable examples of its application to the journals of Economics.  

We pay particular attention to methods that summarize the uncertainty in a ranking produced through the Stigler model-based approach. Our focus on properly accounting for `model-based uncertainty in making comparisons' is close in spirit to \cite{goldstein_96}.  We propose to fit the Stigler model with the quasi-likelihood method \citep{wedderburn_74} to account for inter-dependence among the citations exchanged between pairs of journals, and to summarize estimation uncertainty by using quasi-variances \citep{firth_05b}. We also suggest the use of the ranking lasso penalty \citep{masarotto_12} when fitting the Stigler model, in order to combine the benefits of shrinkage with an enhanced interpretation arising from automatic presentational grouping of journals with similar merits.

The paper is organised as follows. Section \ref{sect:data} describes the data collected from the Web of Science database compiled by Thomson Reuters; then as preliminary background to the paper's main content on journal rankings, Section \ref{sect:cluster} illustrates the use of cluster analysis to identify groups of Statistics journals sharing similar aims and types of content. Section \ref{sect:ranking} provides a  brief summary of journal rankings published by Thomson Reuters in the Journal Citation Reports. Section \ref{sect:stigler} discusses the Stigler method and applies it to the table of cross-citations among Statistics journals. 
Section \ref{sect:RAE} compares journal ratings based on citation data with results from the UK Research Assessment Exercise, and Section \ref{sect:conclusions} collects some concluding remarks.

The citation data set and the computer code used for the analyses  written in the \texttt{R} language \citep{R_11} are made available in the Supplementary Web Materials.  

\section{The Web of Science database}\label{sect:data}

The database used for our analyses is the 2010 edition of the Web of Science produced by Thomson Reuters. The citation data contained in the database are employed to compile the Journal Citation Reports (JCR), whose Science Edition summarizes citation exchange among more than 8,000 journals in science and technology. Within the JCR, scholarly journals are grouped into 171 overlapping subject categories. In particular, in 2010 the \emph{Statistics and Probability} category comprised 110 journals.  The choice of the journals that are encompassed in this category is to some extent arbitrary. The Scopus database, which is the main commercial competitor of Web of Science,  in 2010 included in its Statistics and Probability category 105 journals, but only about two thirds of them were classified in the same category within Web of Science. The Statistics and Probability category contains also journals related to fields such as Econometrics, Chemistry, Computational Biology, Engineering and Psychometrics.

A severe criticism of the Impact Factor relates to the time period used for its calculation. The standard version of the Impact Factor considers citations received to articles published in the previous two years. This period is too short to reach the peak of citations of an article, especially in mathematical disciplines \citep{hall_09}. \citet{nierop_09} finds that articles published in Statistics journals typically reach the peak of their citations more than three years after publication; as reported by the JCR, the median age of the articles cited in this category is more than 10 years. Thomson Reuters acknowledges this issue and computes a second version of the Impact Factor using citations to papers published in the previous five years. Recent published alternatives to the Impact Factor, to be discussed in Section \ref{sect:ranking}, also count citations to articles that appeared in the previous five years. The present paper considers citations of articles published in the previous ten years, in order to capture the influence, over a more substantial period, of work published in statistical journals. 

A key requirement for the methods described here, as well as in our view for any sensible analysis of citation data, is that the journals jointly analysed should be as homogeneous as possible. 
Accordingly, analyses are conducted on a subset of the journals from the Statistics and Probability category, among which there is a relatively high level of citation exchange. The selection is obtained by discarding journals in Probability, Econometrics, Computational Biology, Chemometrics and Engineering, and other journals not sufficiently related to the majority of the journals in the selection. Furthermore, journals recently established, and thus lacking a record of ten years of citable items, also are dropped. The final selection consists of the 47 journals listed in Table \ref{tab:journals}. Obviously, the methods discussed in this paper can be similarly applied to other selections motivated by different purposes. For example, a statistician interested in applications to Economics might consider a different selection with journals of econometrics and statistical methodology, discarding instead journals oriented towards biomedical applications.

\begin{table}
\caption{\label{tab:journals}List of selected Statistics journals, with abbreviations used in the paper.}
\centering
\fbox{%
\begin{tabular}{ll}
Journal name & Abbreviation \\ 
  \hline\hline
American Statistician & AmS \\ 
  Annals of Statistics & AoS \\ 
  Annals of the Institute of Statistical Mathematics & AISM \\ 
  Australian and New Zealand Journal of Statistics & ANZS \\ 
  Bernoulli & Bern \\ 
  Biometrical Journal & BioJ \\ 
  Biometrics & Bcs \\ 
  Biometrika & Bka \\ 
  Biostatistics & Biost \\ 
  Canadian Journal of Statistics & CJS \\ 
  Communications in Statistics - Simulation and Computation & CSSC \\ 
  Communications in Statistics - Theory and Methods & CSTM \\ 
  Computational Statistics & CmpSt \\ 
  Computational Statistics and Data Analysis & CSDA \\ 
  Environmental and Ecological Statistics & EES \\ 
  Environmetrics & Envr \\ 
  International Statistical Review & ISR \\ 
  Journal of Agricultural, Biological and Environmental Statistics  & JABES \\ 
  Journal of Applied Statistics & JAS \\ 
  Journal of Biopharmaceutical Statistics & JBS \\ 
  Journal of Computational and Graphical Statistics & JCGS \\ 
  Journal of Multivariate Analysis & JMA \\ 
  Journal of Nonparametric Statistics & JNS \\ 
  Journal of Statistical Computation and Simulation & JSCS \\ 
  Journal of Statistical Planning and Inference & JSPI \\ 
  Journal of Statistical Software & JSS \\ 
  Journal of the American Statistical Association & JASA \\ 
  Journal of the Royal Statistical Society Series A & JRSS-A \\ 
  Journal of the Royal Statistical Society Series B & JRSS-B \\ 
  Journal of the Royal Statistical Society Series C & JRSS-C \\ 
  Journal of Time Series Analysis & JTSA \\ 
  Lifetime Data Analysis & LDA \\ 
  Metrika & Mtka \\ 
  Scandinavian Journal of Statistics & SJS \\ 
  Stata Journal & StataJ \\ 
  Statistica Neerlandica & StNee \\ 
  Statistica Sinica & StSin \\ 
  Statistical Methods in Medical Research & SMMR \\ 
  Statistical Modelling & StMod \\ 
  Statistical Papers & StPap \\ 
  Statistical Science & StSci \\ 
  Statistics    & Stats \\ 
  Statistics and Computing & StCmp \\ 
  Statistics and Probability Letters & SPL \\ 
  Statistics in Medicine & StMed \\ 
  Technometrics & Tech \\ 
  Test & Test \\ 
\end{tabular}}
\end{table}

The JCR database supplies detailed information about the citations exchanged between pairs of journals through the \emph{Cited Journal Table} and the \emph{Citing Journal Table}. The Cited Journal Table for journal $i$ contains the number of times that articles published in journal $j$ during 2010 cite articles published in journal $i$ in previous years. Similarly, the Citing Journal Table for journal $i$ contains the number of times that articles published in journal $j$ in previous years were cited in journal $i$ during 2010\null. Both of the tables contain some very modest loss of information. In fact, all journals that cite journal $i$ are listed in the Cited Journal Table for journal $i$ only if the number of citing journals is less than $25$. Otherwise, the Cited Journal Table reports only those journals that cite journal $i$ at least twice in \emph{all past years},  thus counting also citations to papers published earlier than the decade 2001--2010 here considered. Remaining journals that cite journal $i$ only once in all past years are collected in the category `all others'. Information on journals cited only once is similarly treated in the Citing Journal Table. 

Cited and Citing Journal Tables allow construction of the cross-citation matrix ${\bf C}=[ c_{ij}]$, where $c_{ij}$ is the number of citations from articles published in journal $j$ in 2010 to papers published in journal $i$ in the chosen time window ($i=1,\ldots, n$)\null. In our analyses, $n=47$, the number of selected Statistics journals, and the time window is the previous ten years. In the rest of this section we provide summary information about citations made and received by each Statistics journal at aggregate level, while Sections \ref{sect:cluster} and \ref{sect:stigler} discuss statistical analyses derived from citations exchanged by pairs of journals. 

\begin{figure}
\centering
\makebox{\includegraphics[scale=.55]{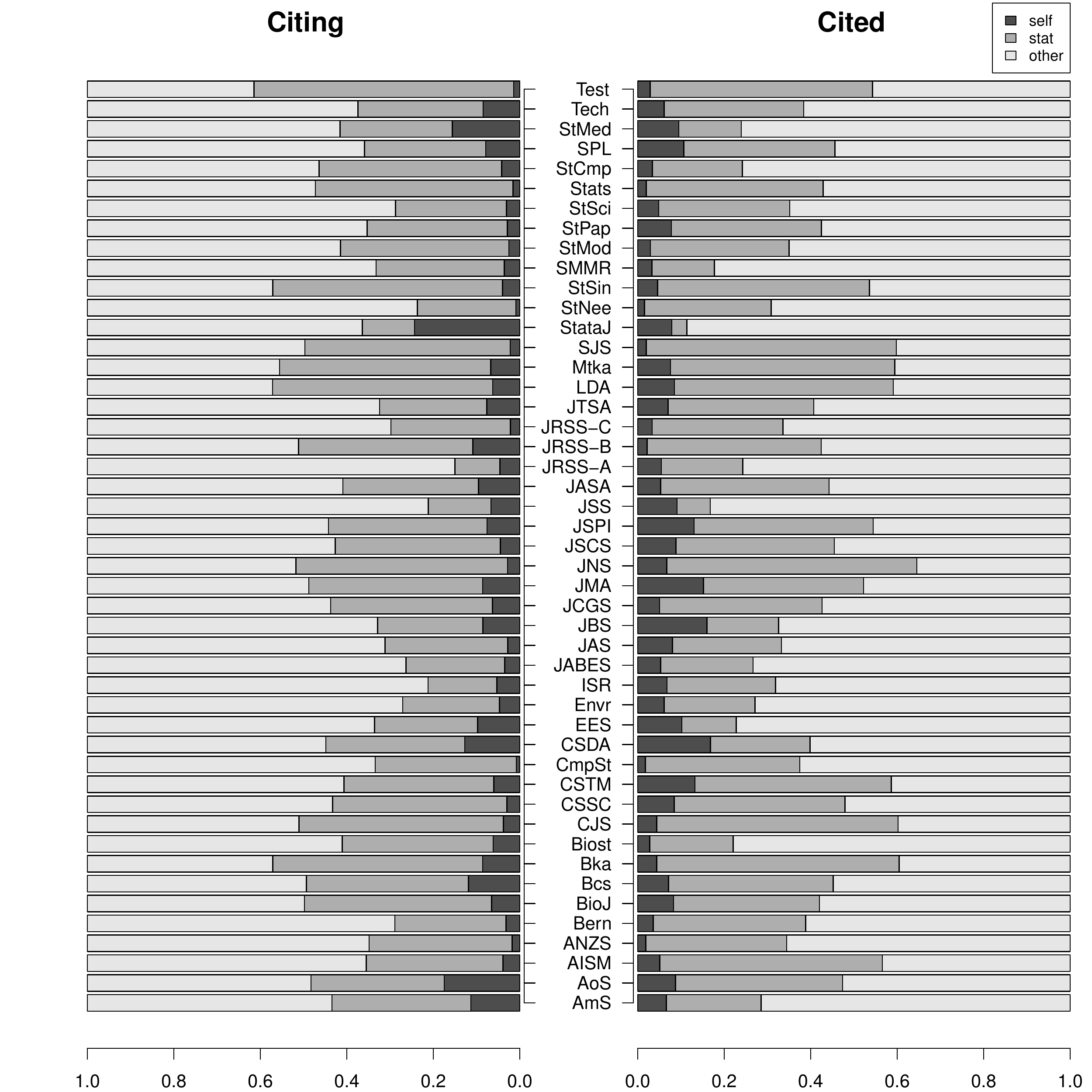}}
\caption{\label{fig:cited}Bar plots of citations made (\texttt{Citing}, left panel) and received (\texttt{Cited}, right panel) for the selected Statistics journals, as listed in Table \ref{tab:descriptive}, based on Journal Citation Reports  2010. For each journal, the bar displays the proportion of self-citations (dark grey), the proportion of citations made/received to/from other Statistics journals in the list (mid grey), and to/from journals not in the list (light grey)\null.}
\end{figure}

Table \ref{tab:descriptive} shows the citations made by papers published in each Statistics journal in 2010 to papers published in other  journals  in the decade 2001--2010, as well as the citations that the papers published in each Statistics journal in   2001--2010 received from papers published in other journals in 2010. The same information is visualized in the back-to-back bar plots of Figure \ref{fig:cited}.  Citations made and received are classified into three categories, namely journal self-citations from a paper published in a journal to another paper in the same journal, citations to/from journals in the list of selected Statistics journals, and citations to/from journals not in the selection.  

The total numbers of citations reported in the second and fifth columns of Table \ref{tab:descriptive} include citations given or received by all journals included in the Web of Science database, not only those in the field of Statistics. The totals are influenced by journals' sizes and by the citation patterns of other categories to which journals are related. The number of references to articles published in 2001--2010 ranges from 275 for citations made in \emph{Statistical Modelling}, which has a small size publishing around 350--400 pages per year, to 4,022 for \emph{Statistics in Medicine}, a large journal with size ranging from 3,500 to 6,000 pages annually in the period examined.
The number of citations from a journal to articles in the same journal is quite variable and ranges from 0.8\% of all citations for \emph{Computational Statistics} to 24\% for \emph{Stata Journal}. On average, 6\% of the references in a journal are to articles appearing in the same journal and 40\% of references are addressed to journals in the list, including journal self-citations. The \emph{Journal of the Royal Statistical Society} Series A has the lowest percentage of citations to {other} journals in the list, at only 10\%. Had we kept the whole \emph{Statistics and Probability} category of the JCR,  that percentage would have risen by just 2 points to 12\%; most of the references appearing in \emph{Journal of the Royal Statistical Society} Series A are to journals outside the Statistics and Probability category.

The number of citations received ranges from 168 for \emph{Computational Statistics} to 6,602 for \emph{Statistics in Medicine}. Clearly, the numbers are influenced by the size of the journal. For example, the small number of citations received by \emph{Computational Statistics} relates to only around 700 pages published per year by that journal. The citations received are influenced also by the citation patterns of other subject categories. In particular, the number of citations received by a journal oriented towards medical applications benefits from communication with a large field including many high-impact journals.  For example, around 75\% of the citations received by \emph{Statistics in Medicine} came from journals outside the list of Statistics journals, mostly from medical journals. On average, 7\% of the citations received by journals in the list came from the same journal and 40\% were from journals in the list.   

As stated already, the Statistics journals upon which we focus have been selected from the Statistics and Probability category of the JCR, with the aim of retaining those which communicate more. The median fraction of citations from journals discarded from our selection to journals in the list is only 4\%, while the median fraction of citations received by non-selected journals from journals in the list is 7\%. An important example of an excluded journal is \emph{Econometrica}, which was ranked in leading positions by all of the published citation indices. \emph{Econometrica} had only about 2\% of its references addressed to other journals in our list, and received only 5\% of its citations from journals within our list.

\begin{table}
\caption{\label{tab:descriptive}Citations made (\texttt{Citing}) and received (\texttt{Cited}) in 2010 to/from articles published in 2001-2010. Columns are total citations (\texttt{Total}), proportion of citations that are journal self-citations (\texttt{Self}), and proportion of citations that are to/from Statistics journals (\texttt{Stat.}), including journal self-citations. Journal abbreviations are as in Table \ref{tab:journals}.}
\centering
\fbox{%
\begin{tabular}{lrrrrrrrr}
  \hline
   \hspace{.1cm} & \multicolumn{3}{c}{Citing} && \multicolumn{3}{c}{Cited} & \\
\cline{2-4}  \cline{6-8}
 Journal & Total & Self   & Stat. && Total & Self   & Stat.  & \\ 
  \hline\hline
AmS & 380 & 0.11 & 0.43 && 648 & 0.07 & 0.29 \\ 
  AoS & 1663 & 0.17 & 0.48 && 3335 & 0.09 & 0.47 \\ 
  AISM & 459 & 0.04 & 0.36 && 350 & 0.05 & 0.57 \\ 
  ANZS & 284 & 0.02 & 0.35 && 270 & 0.02 & 0.34 \\ 
  Bern & 692 & 0.03 & 0.29 && 615 & 0.04 & 0.39 \\ 
  BioJ & 845 & 0.07 & 0.50 && 664 & 0.08 & 0.42 \\ 
  Bcs & 1606 & 0.12 & 0.49 && 2669 & 0.07 & 0.45 \\ 
  Bka & 872 & 0.09 & 0.57 && 1713 & 0.04 & 0.60 \\ 
  Biost & 874 & 0.06 & 0.41 && 1948 & 0.03 & 0.22 \\ 
  CJS & 419 & 0.04 & 0.51 && 362 & 0.04 & 0.60 \\ 
  CSSC & 966 & 0.03 & 0.43 && 344 & 0.08 & 0.48 \\ 
  CSTM & 1580 & 0.06 & 0.41 && 718 & 0.13 & 0.59 \\ 
  CmpSt & 371 & 0.01 & 0.33 && 168 & 0.02 & 0.38 \\ 
  CSDA & 3820 & 0.13 & 0.45 && 2891 & 0.17 & 0.40 \\ 
  EES & 399 & 0.10 & 0.34 && 382 & 0.10 & 0.23 \\ 
  Envr & 657 & 0.05 & 0.27 && 505 & 0.06 & 0.27 \\ 
  ISR & 377 & 0.05 & 0.21 && 295 & 0.07 & 0.32 \\ 
  JABES & 456 & 0.04 & 0.26 && 300 & 0.05 & 0.27 \\ 
  JAS & 1248 & 0.03 & 0.31 && 436 & 0.08 & 0.33 \\ 
  JBS & 1132 & 0.09 & 0.33 && 605 & 0.16 & 0.33 \\ 
  JCGS & 697 & 0.06 & 0.44 && 870 & 0.05 & 0.43 \\ 
  JMA & 2167 & 0.09 & 0.49 && 1225 & 0.15 & 0.52 \\ 
  JNS & 562 & 0.03 & 0.52 && 237 & 0.07 & 0.65 \\ 
  JSCS & 736 & 0.04 & 0.43 && 374 & 0.09 & 0.45 \\ 
  JSPI & 3019 & 0.08 & 0.44 && 1756 & 0.13 & 0.54 \\ 
  JSS & 1361 & 0.07 & 0.21 && 1001 & 0.09 & 0.17 \\ 
  JASA & 2434 & 0.10 & 0.41 && 4389 & 0.05 & 0.44 \\ 
  JRSS-A & 852 & 0.05 & 0.15 && 716 & 0.05 & 0.24 \\ 
  JRSS-B & 506 & 0.11 & 0.51 && 2554 & 0.02 & 0.42 \\ 
  JRSS-C & 731 & 0.02 & 0.30 && 479 & 0.03 & 0.34 \\ 
  JTSA & 327 & 0.08 & 0.32 && 356 & 0.07 & 0.41 \\ 
  LDA & 334 & 0.06 & 0.57 && 247 & 0.09 & 0.59 \\ 
  Mtka & 297 & 0.07 & 0.56 && 264 & 0.08 & 0.59 \\ 
  SJS & 493 & 0.02 & 0.50 && 562 & 0.02 & 0.60 \\ 
  StataJ & 316 & 0.24 & 0.36 && 977 & 0.08 & 0.11 \\ 
  StNee & 325 & 0.01 & 0.24 && 191 & 0.02 & 0.31 \\ 
  StSin & 1070 & 0.04 & 0.57 && 935 & 0.05 & 0.54 \\ 
  SMMR & 746 & 0.04 & 0.33 && 813 & 0.03 & 0.18 \\ 
  StMod & 275 & 0.03 & 0.41 && 237 & 0.03 & 0.35 \\ 
  StPap & 518 & 0.03 & 0.35 && 193 & 0.08 & 0.42 \\ 
  StSci & 1454 & 0.03 & 0.29 && 924 & 0.05 & 0.35 \\ 
  Stats & 311 & 0.02 & 0.47 && 254 & 0.02 & 0.43 \\ 
  StCmp & 575 & 0.04 & 0.46 && 710 & 0.03 & 0.24 \\ 
  SPL & 1828 & 0.08 & 0.36 && 1348 & 0.11 & 0.46 \\ 
  StMed & 4022 & 0.16 & 0.42 && 6602 & 0.10 & 0.24 \\ 
  Tech & 494 & 0.09 & 0.37 && 688 & 0.06 & 0.38 \\ 
  Test & 498 & 0.01 & 0.61 && 243 & 0.03 & 0.54 \\ 
\end{tabular}}
\end{table}

\section{Clustering journals}\label{sect:cluster}

Statistics journals have different stated objectives, and different types of content. Some journals emphasize applications and modelling, while others focus on theoretical and mathematical developments, or deal with computational and algorithmic aspects of statistical analysis. Applied journals are often targeted to particular areas, such as, for example, statistics for medical applications, or for environmental sciences. Therefore, it is quite natural to consider whether the cross-citation matrix ${\bf C}$ allows the identification of groups of journals with similar aims and types of content. Clustering of scholarly journals has been extensively discussed in the bibliometric literature and a variety of clustering methods have been considered. Examples include the hill-climbing method \citep{carpenter_73}, $k$-means \citep{boyack_05}, and methods based on graph theory \citep{leydesdorff_04, liu_12}.

\begin{figure}
\centering
\makebox{\includegraphics[scale=.52]{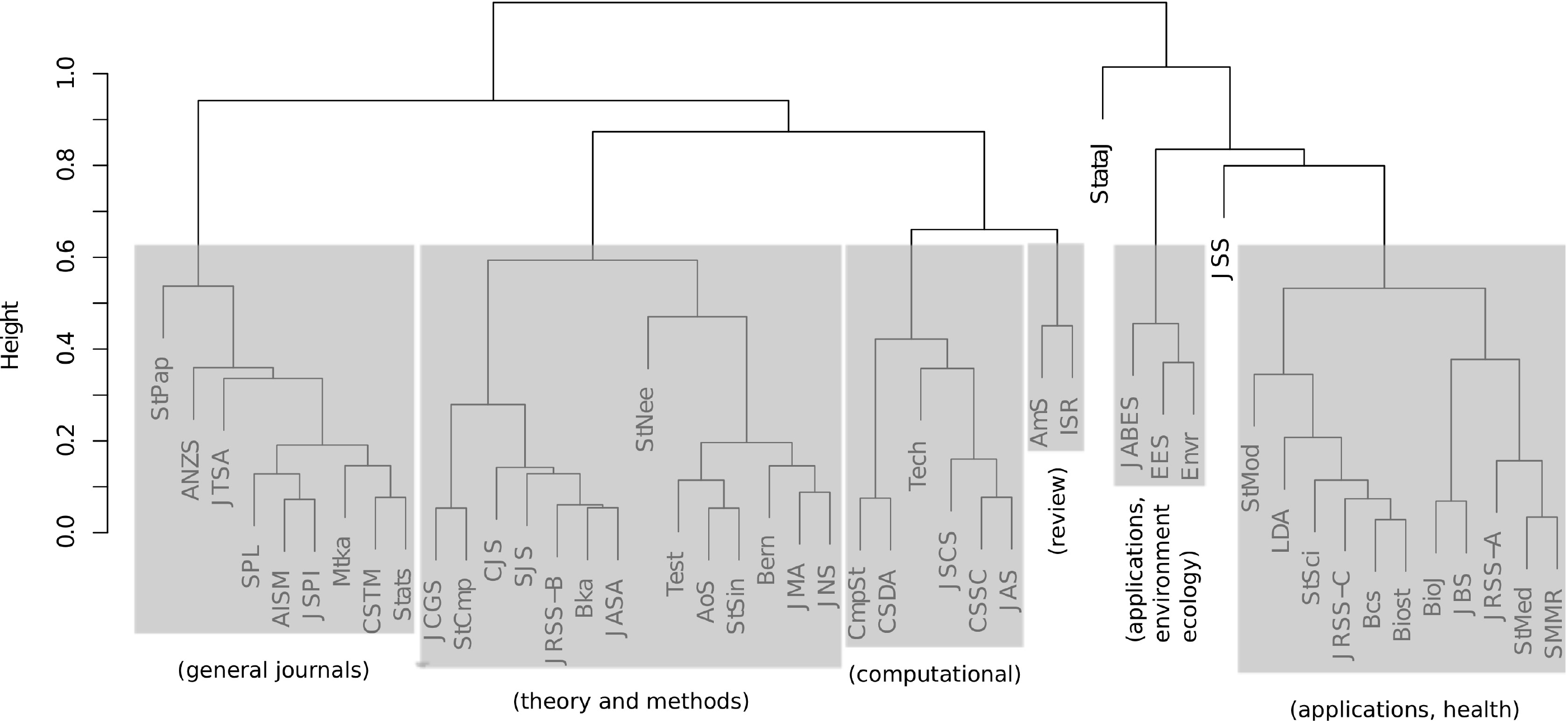}}
\caption{\label{fig:dendro}Dendrogram of complete linkage hierarchical cluster analysis. Clusters obtained by cutting the dendrogram at height 0.6 are identified by grey boxes.}
\end{figure}

Consider the total number $t_{ij}$ of citations exchanged between  journals $i$ and $j$, 
\begin{equation}\label{eq:totals}
t_{ij}=
\begin{cases}
c_{ij}+c_{ji}, & \text{for } i\neq j \\
c_{ii}, & \text{for } i=j.
\end{cases}
\end{equation}
Among various possibilities --- see, for example, \cite{boyack_05} --- the distance between two journals can be measured by quantity $d_{ij}=1-\rho_{ij}$, where $\rho_{ij}$ is the Pearson correlation coefficient of variables $t_{ik}$ and $t_{jk}$ ($k=1,\ldots,n$), \emph{i.e.},  
$$
\rho_{ij}=\frac{\sum_{k=1}^{n}\left(t_{ik}- \bar{t}_{i}\right)\left(t_{jk}-\bar{t}_j\right)}{\sqrt{\sum_{k=1}^{n} \left(t_{ik}-\bar{t}_i\right)^{2}\sum_{k=1}^{n} \left(t_{jk}-\bar{t}_{j}\right)^{2}}},
$$
with $\bar{t}_{i}=\sum_{k=1}^{ n} t_{ik}/n$. Among the many available clustering algorithms,  we consider a hierarchical agglomerative cluster analysis with complete linkage \citep{kaufman_90}.
The clustering process is visualized through the dendrogram in Figure \ref{fig:dendro}. Visual inspection of the dendrogram suggests cutting it at height $0.6,$ thereby obtaining eight clusters, two of which are singletons. The identified clusters are grouped in grey boxes in Figure \ref{fig:dendro}. 

We comment first on the groups and later on the singletons, following the order of the journals in the dendrogram from left to right.  The first group includes a large number of general journals concerned with theory and methods of Statistics, but also with applications. Among others, the group includes  \emph{Journal of Time Series Analysis}, \emph{Journal of Statistical Planning and Inference}, and \emph{Annals of the Institute of Statistical Mathematics}.  

The second group contains the leading journals in the development of statistical theory and methods: \emph{Annals of Statistics}, \emph{Biometrika}, \emph{Journal of the American Statistical Association} and \emph{Journal of the Royal Statistical Society} Series B\null. The group includes also other methodological journals such as \emph{Bernoulli}, \emph{Scandinavian Journal of Statistics} and \emph{Statistica Sinica}\null. It is possible to identify some natural subgroups: \emph{Journal of Computational and Graphical Statistics} and \emph{Statistics and Computing}; \emph{Biometrika}, \emph{Journal of the Royal Statistical Society} Series B, and \emph{Journal of the American Statistical Association}; \emph{Annals of Statistics} and \emph{Statistica Sinica}.  

The third group comprises journals mostly dealing with computational aspects of Statistics, such as  \emph{Computational Statistics and Data Analysis}, \emph{Communications in Statistics -- Simulation and Computation}, \emph{Computational Statistics}, and \emph{Journal of Statistical Computation and Simulation}. Other members of the group with a less direct orientation towards computational methods are \emph{Technometrics} and \emph{Journal of Applied Statistics}. 

The fourth group includes just two journals both of which publish mainly review articles, namely \emph{American Statistician} and \emph{International Statistical Review}.

The fifth group comprises the three journals specializing in ecological and environmental applications: \emph{Journal of Agricultural, Biological and Environmental Statistics}, \emph{Environmental and Ecological Statistics} and \emph{Environmetrics}. 
 
The last group includes various journals emphasising applications, especially to health sciences and similar areas.  It encompasses journals oriented towards biological and medical applications such as \emph{Biometrics} and \emph{Statistics in Medicine}, and also journals publishing papers about more general statistical applications, such as \emph{Journal of the Royal Statistical Society} Series A and Series C\null. The review journal \emph{Statistical Science} also falls into this group; it is not grouped together with the other two review journals already mentioned.  Within the group there are some natural sub-groupings:  \emph{Statistics in Medicine} with \emph{Statistical Methods in Medical Research}; and \emph{Biometrics} with \emph{Biostatistics}.   
 
Finally, and perhaps not surprisingly, the two singletons are the software-oriented \emph{Journal of Statistical Software} and \emph{Stata Journal}. The latter is, by some distance, the most remote journal in the list according to the measure of distance used here.  

\section{Ranking journals}\label{sect:ranking}


The Thomson Reuters JCR website annually publishes various rating indices, the best known being the already mentioned Impact Factor. Thomson Reuters also publishes the \emph{Immediacy Index}, which describes the average number of times an article is cited in the year of its publication. The Immediacy Index is unsuitable for evaluating Statistics journals, but it could be worthy of attention in fields where citations occur very quickly, for example some areas of neuroscience and other life sciences.

It is well known in the bibliometric literature that the calculation of the Impact Factor contains some important inconsistencies \citep{glanzel_2002}. The numerator of the Impact Factor includes citations to all items, while the number of citable items in the denominator excludes letters to the editor and editorials; such letters are an important element of some journals, notably medical journals. The inclusion of self-citations, defined as citations from a journal to articles in the same journal, exposes the Impact Factor to possible manipulation by editors. Indeed, \citet{sevinc_04}, \citet{frandsen_07} and \citet{wilhite_12} report instances where authors were asked to add irrelevant references to their articles, presumably with the aim of increasing the Impact Factor of the journal. As previously mentioned, recently Thomson Reuters has made available also the Impact Factor without journal self-citations. 
Journal self-citations can also be a consequence of authors' preferring to cite papers published in the same journal instead of equally relevant papers published elsewhere, particularly if they perceive such self-citation as likely to be welcomed by the journal's editors. Nevertheless, the potential for such behaviour should not lead to the conclusion that self-citations are always unfair. Many self-citations are likely to be genuine, especially since scholars often select a journal for submission of their work according to the presence of previously published papers on related topics. 

The \emph{Eigenfactor Score} and the derived \emph{Article Influence Score}  \citep{bergstrom_07, west_10} have been proposed to overcome the limitations of the Impact Factor. Both the Eigenfactor  and the Article Influence Score  are computed over a five-year time period, with journal self-citations removed in order to eliminate possible sources of manipulation.
The idea underlying the Eigenfactor Score  is that the importance of a journal relates to the time spent by scholars in reading that journal. As stated by \citet{bergstrom_07}, it is possible to imagine that a scholar starts reading an article selected at random. Then, the scholar randomly selects another article from the references of the first paper and reads it. Afterwards, a further article is selected at random from the references included in the previous one and the process may go on \emph{ad infinitum}. In such a process, the time spent in reading a journal might reasonably be regarded as an indicator of that journal's importance. 

Apart from modifications needed to account for special cases such as journals that do not cite any other journal, the Eigenfactor  algorithm is summarized as follows. The Eigenfactor  is computed from the normalized citation matrix $\tilde{\bf C}=[\tilde{c}_{ij}]$,  whose elements are the citations $c_{ij}$ from journal $j$ to articles published in the previous five years in journal $i$ divided by the total number of references in $j$ in those years, $\tilde{c}_{ij}=c_{ij}/\sum_{i=1}^n c_{ij}$.  The diagonal elements of $\tilde{\bf C}$ are set to zero, to discard self-citations.  A further ingredient of the Eigenfactor  is the vector of normalized numbers of articles  $\bm{a}=(a_1, \ldots, a_n)^{\top}$, with $a_i$ being the number of articles published by journal $i$ during the five-year period divided by the number of articles published by all considered journals. Let $\bm{e}^{\top}$ be the row vector of ones, so that $\bm{a}\bm{e}^{\top}$ is a matrix with all identical columns $\bm{a}$. Then 
$$
{\bf P}=\lambda \tilde{{\bf C}}+(1-\lambda) \bm{a} \bm{e}^{\top}
$$
is the transition matrix of a Markov process that assigns probability $\lambda$ to a random movement in the journal citation network, and probability $1-\lambda$ to a random jump to any journal; for jumps of the latter kind, destination-journal attractiveness is simply proportional to size. 

The damping parameter $\lambda$ is set to $0.85$, just as in the \emph{PageRank}  algorithm  at the basis of the Google search engine; see \cite{brin_98}. The leading eigenvector $\bm{\psi}$ of ${\bf P}$ corresponds to the  steady-state fraction of time spent reading each journal. The Eigenfactor Score  $\text{EF}_i$ for journal $i$ is defined as `the percentage of the total weighted citations that journal $i$ receives'; that is,
$$
\text{EF}_i=100 \frac{ [\tilde{{\bf C}} \bm{\psi}]_{i}}{\sum_{i=1}^n [\tilde{{\bf C}} \bm{\psi}]_{i}}, \quad i=1, \ldots, n,
$$
where $[{\bf x}]_i$ denotes the $i$th element of vector ${\bf x}$\null. See \url{www. eigenfactor.org/methods.pdf} for more details of the methodology behind the Eigenfactor  algorithm.

The Eigenfactor  `measures the total influence of a journal on the scholarly literature' \citep{bergstrom_07} and thus it depends on the number of articles published by a journal. The Article Influence Score  $\text{AI}_{i}$ of journal $i$ is instead a measure of the per-article citation influence of the journal, obtained by normalizing the Eigenfactor as follows: 
$$
\text{AI}_{i}=0.01\frac{\text{EF}_{i}}{a_{i}}, \quad i=1, \ldots, n.
$$
Distinctive aspects of the Article Influence Score with respect to the Impact Factor are:
\begin{enumerate}
\item The use of a formal stochastic model to derive the journal ranking;
\item The use of bivariate data --- the cross-citations $c_{ij}$ --- in contrast to the univariate citation counts used by the Impact Factor.
\end{enumerate}
An appealing feature of the Article Influence Score  is that citations are weighted according to the importance of the source, whereas the Impact Factor counts all citations equally \citep{franceschet_10}\null. Accordingly, the bibliometric literature classifies  the Article Influence Score as a measure of journal `prestige' and the Impact Factor as a measure of journal `popularity'  \citep{bollen_06}. 
Table \ref{tab:characteristics} summarizes some of the main features of the ranking methods discussed in this section and
also of the Stigler model that will be discussed in Section \ref{sect:stigler} below.

\begin{table}
\caption{\label{tab:characteristics}
Characteristics of the journal rankings derived from
Journal Citation Reports.  Rankings are: Immediacy Index (\texttt{II}),
Impact Factor (\texttt{IF}), Impact Factor without self-citations (\texttt{IFno}),
five-year Impact Factor (\texttt{IF5}), Article
Influence Score (\texttt{AI}), and the Stigler model studied in this paper (\texttt{SM}).
The `Data' column indicates whether the data used are bivariate
cross-citation counts or only univariate citation counts.
`Global/Local' relates to whether a ranking is `local' to the
main journals of Statistics, or `global' in that it is applied
across disciplines.}
\centering
\fbox{%
\begin{tabular}{lccccc}
 &  Citation &  Stochastic  & &  Excludes  &  Global/\\ 
Ranking & Period (yrs) & Model  & Data & Self-citation & Local\\
\hline
\texttt{II} & 1   &  none  & univariate   & no   & global \\
\texttt{IF} & 2   & none & univariate & no   & global \\
\texttt{IFno} & 2    & none & univariate & yes   & global \\
\texttt{IF5} & 5   & none & univariate & no & global\\
\hline 
\texttt{AI} & 5    & Markov & bivariate & yes & global\\
& & process & & & \\
\hline 
\texttt{SM} & 10  & Bradley-& bivariate & yes & local \\
& &  Terry& & & 
\end{tabular}
}
\end{table}

The rankings of the selected Statistics journals according to Impact Factor, Impact Factor without journal self-citations, five-year Impact Factor, Immediacy Index, and Article Influence Score  are reported in columns two to six of Table \ref{tab:rankings}. The substantial variation among those five rankings is the first aspect that leaps to the eye; these different published measures clearly do not yield a common, unambiguous picture of the journals' relative standings.

\begin{table}
\caption{\label{tab:rankings} Rankings of selected Statistics journals based on Journal Citation Reports, 2010 Edition. Columns correspond to Immediacy Index (\texttt{II}),  Impact Factor (\texttt{IF}), Impact Factor without self-citations (\texttt{IFno}), five-year Impact Factor (\texttt{IF5}), Article Influence Score (\texttt{AI}), and the Stigler model (\texttt{SM}). Braces indicate groups identified by the ranking lasso.}
\centering
\fbox{%
\begin{tabular}{r  l  l  l   l   l   l   l}
 Rank & II & IF & IFno & IF5 & AI & SM  & \\ 
  \hline\hline
1 & JSS & JRSS-B & JRSS-B & JRSS-B & JRSS-B & JRSS-B \\ 
  2 & Biost & AoS & Biost & JSS & StSci & AoS \\ 
  3 & SMMR & Biost & AoS & StSci & JASA & Bka \ldelim\}{2}{1mm}[\hspace{5mm}] &\\ 
  4 & StCmp & JSS & JRSS-A & JASA & AoS & JASA \\ 
  5 & AoS & JRSS-A & JSS & Biost & Bka & Bcs \\ 
  6 & EES & StSci & StSci & AoS & Biost & JRSS-A \ldelim\}{6}{1mm}[\hspace{-0.5mm}]&\\ 
  7 & JRSS-B & StMed & StMed & StataJ & StataJ & Bern \\ 
  8 & JCGS & JASA & JASA & SMMR & StCmp & SJS \\ 
  9 & StMed & StataJ & StataJ & JRSS-A & JRSS-A & Biost \\ 
  10 & BioJ & StCmp & StCmp & Bka & JSS & JCGS \\ 
  11 & CSDA & Bka & SMMR & StCmp & Bcs & Tech \\ 
  12 & StSci & SMMR & Bka & StMed & Bern & AmS \ldelim\}{6}{1mm}[\hspace{4mm}]&\\
  13 & JRSS-A & Bcs & EES & Bcs & JCGS & JTSA \\ 
  14 & StSin & EES & Bcs & Tech & SMMR & ISR \\ 
  15 & JBS & Tech & Tech & JCGS & Tech & AISM \\ 
  16 & StataJ & BioJ & BioJ & EES & SJS & CJS \\ 
  17 & Bcs & JCGS & JCGS & CSDA & StMed & StSin \\ 
  18 & Envr & CSDA & Test & SJS & Test & StSci \ldelim\}{13}{1mm}[\hspace{4mm}]&\\
  19 & Bka & JBS & AISM & AmS & CJS & LDA \\ 
  20 & JMA & Test & Bern & JBS & StSin & JRSS-C \\ 
  21 & Tech & JMA & StSin & Bern & JRSS-C & StMed \\ 
  22 & JASA & Bern & LDA & JRSS-C & AmS & ANZS \\ 
  23 & JRSS-C & AmS & JMA & BioJ & JMA & StCmp \\ 
  24 & ISR & AISM & CSDA & JABES & EES & StataJ \\ 
  25 & JNS & StSin & SJS & JMA & JTSA & SPL \\ 
  26 & Test & LDA & ISR & CJS & LDA & StNee \\ 
  27 & Bern & ISR & JBS & Test & BioJ & Envr \\ 
  28 & JABES & SJS & AmS & StMod & StMod & JABES \\ 
  29 & JSPI & Envr & Envr & StSin & CSDA & Mtka \\ 
  30 & SJS & JABES & StMod & LDA & JABES & StMod \\ 
  31 & AmS & StMod & CJS & Envr & AISM & JSPI \ldelim\}{3}{1mm}[\hspace{4.5mm}] &\\ 
  32 & AISM & JSPI & JABES & JTSA & ANZS & SMMR \\ 
  33 & StMod & CJS & JTSA & ISR & ISR & BioJ \\ 
  34 & Mtka & JTSA & JSPI & ANZS & JSPI & JMA \ldelim\}{11}{1mm}[\hspace{4mm}] &\\   
  35 & StNee & JRSS-C & ANZS & JSPI & Envr & EES \\ 
  36 & StPap & ANZS & StPap & AISM & JBS & CSDA \\ 
  37 & SPL & StPap & Mtka & Stats & StNee & JNS \\ 
  38 & ANZS & Mtka & JRSS-C & Mtka & CmpSt & CmpSt \\ 
  39 & LDA & Stats & Stats & CmpSt & JNS & Stats \\ 
  40 & JTSA & CmpSt & CmpSt & StNee & Stats & Test \\ 
  41 & JSCS & JSCS & JSCS & JSCS & Mtka & CSTM \\ 
  42 & CJS & JNS & JNS & StPap & JSCS & JSS \\ 
  43 & CmpSt & SPL & SPL & SPL & StPap & JBS \\ 
  44 & CSTM & CSTM & CSTM & JNS & SPL & JSCS \\ 
  45 & Stats & CSSC & StNee & JAS & CSTM & CSSC \ldelim\}{3}{1mm}[\hspace{3mm}] &\\ 
  46 & JAS & StNee & CSSC & CSTM & CSSC & StPap \\ 
  47 & CSSC & JAS & JAS & CSSC & JAS & JAS \\   
   \end{tabular}}
\end{table}

A diffuse opinion within the statistical community is that the four most prestigious Statistics journals are (in alphabetic order) \emph{Annals of Statistics}, \emph{Biometrika}, \emph{Journal of the American Statistical Association}, and \emph{Journal of the Royal Statistical Society} Series B\null. See, for example, the survey about how statisticians perceive Statistics journals described in \cite{theoharakis_03}\null. Accordingly, a minimal requirement for a ranking of acceptable quality is that the four most prestigious journals should occupy prominent positions. Following this criterion, the least satisfactory ranking is, as expected, the one based on the Immediacy Index, which ranks \emph{Journal of the American Statistical Association} only 22nd and \emph{Biometrika} just a few positions ahead at 19th. 

In the three versions of Impact Factor ranking, \emph{Journal of the Royal Statistical Society} Series B always occupies first position, \emph{Annals of Statistics} ranges between second and sixth, \emph{Journal of the American Statistical Association} between fourth and eighth, and \emph{Biometrika} between tenth and twelfth.  The two software journals have quite high Impact Factors: \emph{Journal of Statistical Software} is ranked between second and fifth by the three different Impact Factor versions, while \emph{Stata Journal} is between seventh and ninth.  Other journals ranked highly according to the Impact Factor measures are \emph{Biostatistics} and \emph{Statistical Science}.  

Among the indices published by Thomson Reuters, the Article Influence Score yields the most satisfactory ranking with respect to the four leading journals mentioned above, all of which stand within the first five positions. 

All of the indices discussed in this section are constructed by using the complete Web of Science database, thus counting citations from journals in other fields as well as citations among Statistics and Probability journals.

\section{The Stigler model}\label{sect:stigler}

\citet{stigler_94} considers the export of intellectual influence from a journal in order to determine its importance. The export of influence is measured through the citations received by the journal. Stigler assumes that the log-odds that journal $i$ exports to journal $j$ rather than vice-versa is equal to the difference of the journals' \emph{export scores},
\begin{equation}\label{eq:stigler}
\text{log-odds}\left(\text{journal $i$ is cited by journal $j$}\right)=\mu_{i}-\mu_{j},
\end{equation}
where $\mu_{i}$ is the export score of journal $i$\null. In Stephen Stigler's words `the larger the export score, the greater the propensity to export intellectual influence'. The Stigler model is an example of the Bradley-Terry model \citep{bradley_52, david_63, agresti_02} for paired comparison data. According to  (\ref{eq:stigler}), the citation counts $c_{ij}$ are realizations of binomial variables $C_{ij}$ with expected value
\begin{equation}\label{eq:quasi-stigler}
\text{E}(C_{ij})=t_{ij}  \pi_{ij},
\end{equation}
where $\pi_{ij}=\exp( \mu_i-\mu_j)/\left\{ 1+ \exp( \mu_i-\mu_j)\right\}$ and $t_{ij}$ is the total number of citations exchanged between  journals $i$ and $j$, as defined in (\ref{eq:totals}).


The Stigler model has some attractive features:
\begin{enumerate}
\item {\it Statistical modelling}. Similarly to the Eigenfactor and the derived Article Influence Score, the Stigler method is based on stochastic modelling of a matrix of cross-citation counts.  The methods differ regarding the modelling perspective --- Markov process for Eigenfactor versus Bradley-Terry model in the Stigler method --- and, perhaps most importantly, the use of formal statistical methods.  The Stigler model is calibrated through well-established statistical fitting methods, such as maximum likelihood or quasi-likelihood (see Section \ref{sect:fitting}), with  estimation uncertainty summarized accordingly  (Section \ref{sect:uncertainty}). Moreover, Stigler-model assumptions are readily checked by the analysis of suitably defined residuals, as described in Section \ref{sect:validation}. 
\item {\it The size of the journals is not important}. Rankings based on the Stigler model are not affected by the numbers of papers published. As shown by \citeauthor{stigler_94} (\citeyear{stigler_94}, pg.~102), if two journals are merged into a single journal then the odds in favour of that `super' journal against any third journal is a weighted average of the odds for the two separate journals against the third one.  Normalization for journal size, which 
is explicit in the definitions of various 
Impact Factor and Article Influence measures, is thus implicit
for the Stigler model. 
\item {\it Journal self-citations are not counted}. In contrast to the standard Impact Factor, rankings based on journal export scores $\mu_i$ are not affected by the risk of manipulation through journal self-citations. 
\item {\it Only citations between journals under comparison are counted}. If the Stigler model is applied to the list of 47 Statistics journals, then only citations among these journals are counted.  Such an application of the Stigler model thus aims unambiguously to measure influence within the research field of Statistics, rather than combining that with potential influence on other research fields. As noted in Table~\ref{tab:characteristics}, this property differentiates the Stigler model from the other ranking indices published by Thomson Reuters, which use citations from all journals in potentially any fields in order to create a `global' ranking of all scholarly journals. 
Obviously it would be possible also to re-compute more `locally' the various Impact Factor measures and/or Eigenfactor-based indices, by using only citations exchanged between the journals in a restricted set to be compared.  
\item {\it Citing journal is taken into account}. Like the 
 Article Influence Score, the Stigler model measures journals' relative prestige, because it is derived from bivariate citation counts and thus takes into account the source of each citation.  The Stigler
model decomposes the cross-citation matrix ${\bf C}$ differently, though;
it can be re-expressed in log-linear form as the 
`quasi symmetry' model, 
\begin{equation}
\label{eqn:quasisymmetry} 
E(C_{ij}) = t_{ij}e^{\alpha_i+ \beta_j} , 
\end{equation}  
in which the export score for journal $i$ is 
$\mu_i = \alpha_i - \beta_i$\null.  
\item {\it Lack-of-fit assessment}. \cite{stigler_95} and \cite{liner_04} observed increasing lack of fit of the Stigler model when additional journals that trade little with those already under comparison are included in the analysis. \cite{ritzberger_08} states bluntly that the Stigler model `suffers from a lack of fit' and dismisses it --- incorrectly, in our view --- for that reason.  We agree instead with \cite{liner_04} who suggest that statistical lack-of-fit assessment is another positive feature of the Stigler model that can be used, for example, to identify groups of journals belonging to different research fields, journals which should perhaps not be ranked together.  Certainly the existence of principled lack-of-fit assessment for the Stigler model should not be a reason to prefer other methods for which no such assessment is available.  
\end{enumerate}
See also Table \ref{tab:characteristics} for a comparison of properties of the ranking methods considered in this paper.

\subsection{Model fitting}\label{sect:fitting}

Maximum likelihood estimation of the vector of journal export scores $\bm{\mu}=(\mu_1, \ldots, \mu_n)^{\top}$ can be obtained through standard software for fitting generalized linear models. Alternatively, specialized software such as the \texttt{R} package \texttt{BradleyTerry2} \citep{turner_11} is available through the CRAN repository. Since the Stigler model is specified through pairwise differences of export scores $\mu_i-\mu_j$, model identification requires a constraint, such as, for example, a `reference journal' constraint $\mu_1=0$ or the sum constraint $\sum_{i=1}^n \mu_i=0$\null. Without loss of generality we use the latter constraint in what follows. 

Standard maximum likelihood estimation of the Stigler model would assume that citation counts $c_{ij}$ are realizations of independent binomial variables $C_{ij}$. Such an assumption is likely to be inappropriate, since research citations are not independent of one another in practice; see \citet{cattelan_12} for a general discussion on handling dependence in paired-comparison modelling.  The presence of dependence between citations can be expected to lead to the well-known phenomenon of overdispersion.  A simple way to deal with overdispersion is provided by the method of quasi-likelihood \citep{wedderburn_74}. Accordingly, we consider a `quasi-Stigler' model, 
\begin{equation}\label{eq:quasi-stigler}
\text{E}(C_{ij})=t_{ij}  \pi_{ij} \quad \text{and} \quad \text{var}(C_{ij})=\phi\, t_{ij}  \pi_{ij}  (1-\pi_{ij}),
\end{equation}
where 
$\phi > 0$ is the dispersion parameter. Let $\bm{c}$ be the vector obtained by stacking all citation counts $c_{ij}$ in some arbitrary order, and let $\bm{t}$ and $\bm{\pi}$ be the corresponding vectors of totals $t_{ij}$ and expected values $\pi_{ij}$, respectively. Then estimates of the export scores are obtained by solving the quasi-likelihood estimating equations
\begin{equation}\label{eq:quasi-likelihood}
{\bf D}^\top {\bf V}^{-1} \left( \bm{c}- \bm{t} \bm{\pi}\right)=\bm{0},
\end{equation}
where $\bf{D}$ is the Jacobian of $\bm{\pi}$ with respect to the export scores $\bm{\mu}$, and $\bf{V}=\bf{V}(\bm{\mu})$ is the diagonal matrix with elements $\text{var}(C_{ij})/\phi$\null. Under the assumed model (\ref{eq:quasi-stigler}), quasi-likelihood estimators are consistent and asymptotically normally distributed with variance-covariance matrix  $\phi \left( {\bf D}^\top {\bf V}^{-1} {\bf D} \right)^{-1}$. The dispersion parameter is usually estimated via the squared Pearson residuals as
\begin{equation}\label{eq:overdispersion}
\hat{\phi}=  \frac{1}{m-n+1} \sum_{i< j}^n \frac{\left(c_{ij}-t_{ij}\hat \pi_{ij}\right)^2}{t_{ij}\hat \pi_{ij}(1-\hat \pi_{ij})} ,
\end{equation}
where $\hat{\bm{\pi}}$ is the vector of estimates $\hat \pi_{ij}=\exp( \hat\mu_i-\hat\mu_j)/\left\{ 1+ \exp( \hat\mu_i-\hat\mu_j)\right\}$, with $\hat \mu_i$ being the quasi-likelihood estimate of the export score $\mu_i$, and $m=\sum_{i < j} 1(t_{ij}>0)$ the number of pairs of journals that exchange citations. 
Well-known properties of quasi-likelihood estimation are robustness against misspecification of the variance matrix $\bf{V}$ and optimality within the class of linear unbiased estimating equations. 

The estimate of the dispersion parameter obtained here, for the model applied to Statistics journal cross-citations between 2001 and 2010, is $\hat\phi=1.76$, indicative of overdispersion. The quasi-likelihood estimated export scores of the Statistics journals are reported in Table \ref{tab:BTrank} and will be discussed later in Section \ref{sect:results}.

\subsection{Model validation}\label{sect:validation}

An essential feature of the Stigler model is that the export score of any journal is a constant. In particular, in model (\ref{eq:stigler}) the export score of journal $i$ is not affected by the identity of the citing journal $j$.  Citations exchanged between journals can be seen as results of contests between opposing journals and the residuals for contests involving journal $i$ should not exhibit any relationship with the corresponding estimated export scores of the `opponent' journals $j$.  With this in mind, we define the \emph{journal residual} $r_i$ for journal $i$ as the standardized regression coefficient derived from the linear regression of Pearson residuals involving journal $i$ on the estimated export scores of the corresponding opponent journals. More precisely, the $i$th journal residual is defined here as
$$
r_i = \frac{\sum_{j=1}^n \hat \mu_j \, r_{ij} }{\sqrt{ \hat{\phi} \, \sum_{j=1}^n \hat \mu_j^2}}, 
$$
where 
$r_{ij}$ is the Pearson residual for citations of $i$ by $j$, 
$$
r_{ij}= \frac{c_{ij}-t_{ij}\hat \pi_{ij}}{\sqrt{t_{ij}\hat \pi_{ij}(1-\hat \pi_{ij})}}. 
$$  
The journal residual $r_i$ indicates the extent to which $i$ performs systematically better than predicted by the model either when the opponent $j$ is strong, as indicated by positive-valued journal
residual for $i$, or when the opponent $j$ is weak, as indicated by a negative-valued journal residual for $i$.  The journal residuals thus provide a basis for useful diagnostics, targeted specifically at readily interpretable departures from the assumed model. 

\begin{figure}
\centering
\makebox{
\includegraphics[scale=.51]{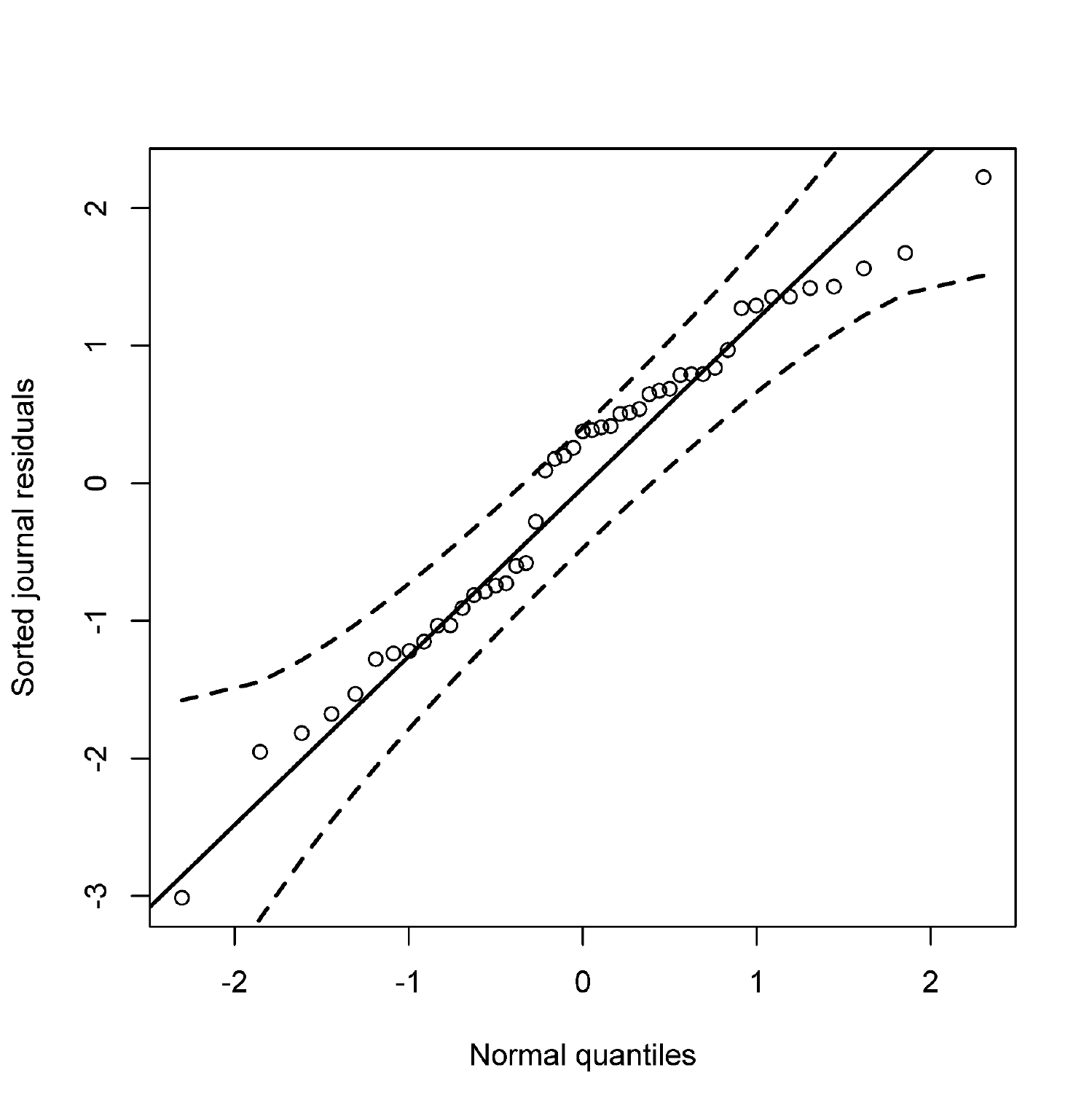} 
\includegraphics[scale=.51]{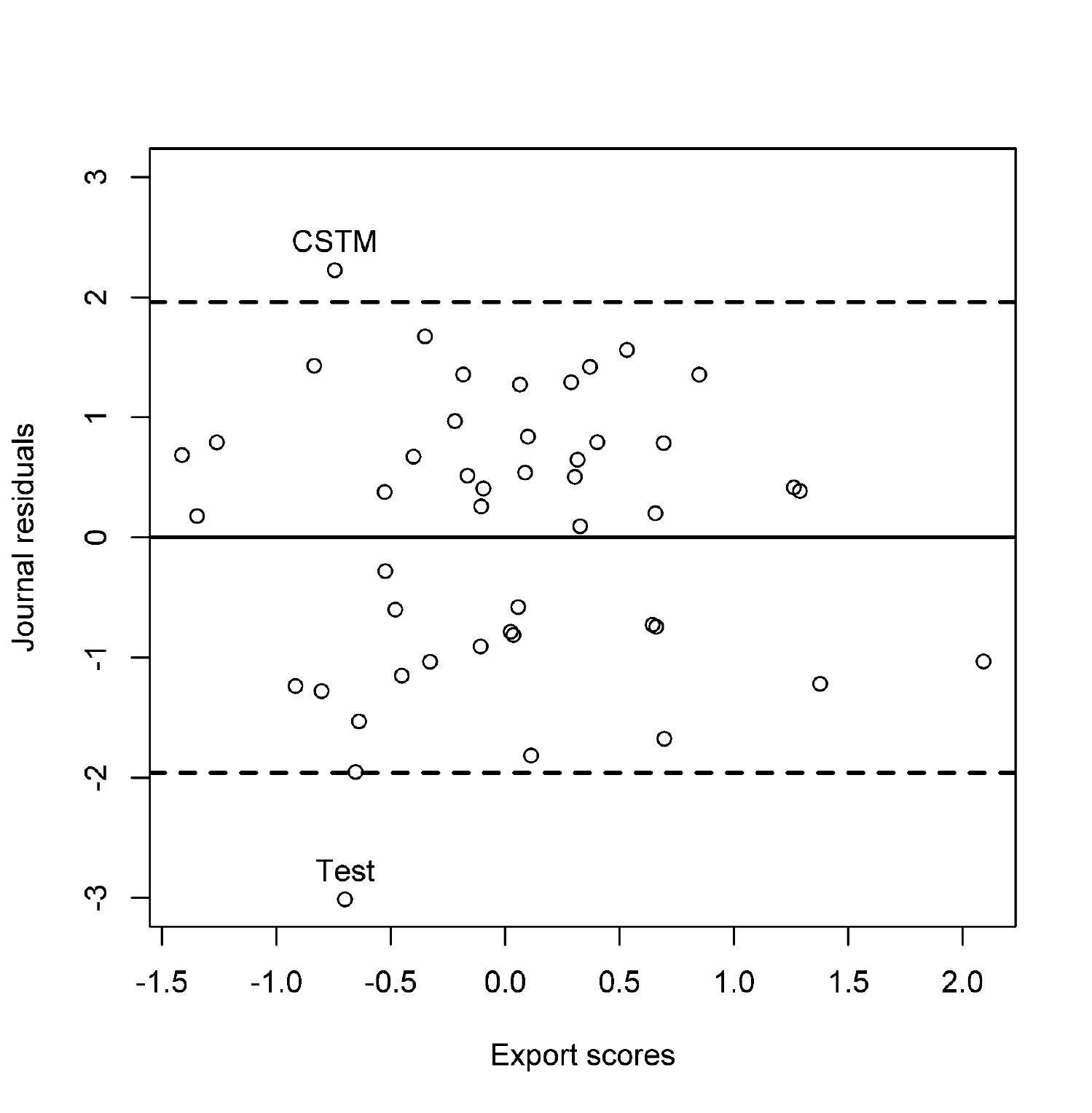}
}
\caption{\label{fig:residuals} Normal probability plot of journal residuals with $95\%$ simulation envelope (left panel) and scatterplot of journal residuals versus estimated journal export scores (right panel).} 
\end{figure}

Under the assumed quasi-Stigler model, journal residuals are approximately realizations of standard normal variables and are unrelated to the export scores. The normal probability plot of the journal residuals displayed in the left panel of Figure \ref{fig:residuals} indicates that the normality assumption is indeed approximately satisfied. The scatterplot of the journal residuals $r_i$ against estimated export scores $\hat\mu_i$ shows no clear pattern; there is no evidence of correlation between journal residuals and export scores. As expected based on approximate normality of the residuals, only two journals --- \emph{i.e.}, $4.3\%$ of journals --- have residuals larger in absolute value than $1.96$. These journals are {\it Communications in Statistics - Theory and Methods} ($r_{\text{CSTM}}=2.23$) and {\it Test} ($r_{\text{Test}}=-3.01$).  The overall conclusion from this graphical inspection of journal residuals is that the assumptions of the quasi-Stigler model appear to be essentially satisfied for the data used here. 

\subsection{Estimation uncertainty}\label{sect:uncertainty}

Estimation uncertainty is commonly unexplored, and is rarely reported, in relation to the various published journal rankings. Despite this lacuna, many academics have produced vibrant critiques of `statistical citation analyses', although such analyses are actually rather non-statistical. Recent research in the bibliometric field has suggested that uncertainty in estimated journal ratings might be estimated via bootstrap simulation; see the already mentioned  \cite{chen_14} and the `stability intervals' for the SNIP index.  A key advantage of the Stigler model over other ranking methods is straightforward quantification of the uncertainty in journal export scores. 

Since the Stigler model is identified through pairwise differences, uncertainty quantification requires the complete variance matrix of  $\hat{\bm{\mu}}$\null. Routine reporting of such a large variance matrix is impracticable for space reasons.  A neat solution is provided through the presentational device of quasi-variances \citep{firth_05b}, constructed in such a way as to allow approximate calculation of any variance of a difference, $\text{var}(\hat \mu_i-\hat \mu_j)$, as if $\hat\mu_i$ and $\hat\mu_j$ were independent:
$$
\text{var}\left( \hat \mu_i -\hat \mu_j \right)\simeq \text{qvar}_i + \text{qvar}_j, \quad \text{for all choices of $i$ and $j$}.
$$
Reporting the estimated export scores with their quasi-variances, then, is an economical way to allow approximate inference on the significance of the difference between any two journals' export scores. The quasi-variances are computed by minimizing a suitable penalty function of the differences between the true variances, $\text{var}\left( \hat \mu_i -\hat \mu_j \right)$, and their quasi-variance representations $\text{qvar}_i+\text{qvar}_j$. See \cite{firth_05b} for details. 

Table \ref{tab:BTrank} reports the estimated journal export scores computed under the sum constraint $\sum_{i=1}^n \mu_i=0$ and the corresponding quasi standard errors, defined as the square root of the quasi-variances. Quasi-variances are calculated by using the \texttt{R}  package \texttt{qvcalc} \citep{firth_10}. For illustration, consider testing whether the export score of \emph{Biometrika} is significantly different from that of the \emph{Journal of the American Statistical Association}. The $z$ test statistic as approximated through the quasi-variances is
$$
z\simeq\frac{\hat \mu_{\text{Bka}}-\hat \mu_{\text{JASA}}}{\sqrt{\text{qvar}_{\text{Bka}}+\text{qvar}_{\text{JASA}}}}=
\frac{1.29-1.26}{\sqrt{0.08^2+0.06^2}}= 0.30.
$$
The `usual'
variances for those two export scores in the sum-constrained parameterization are respectively 0.0376 and 0.0344, and the covariance is 0.0312; thus the `exact' value of the $z$ statistic in this example is 
$$
z=\frac{1.29-1.26}{\sqrt{0.0376-2 \, (0.0312)+ 0.0344 }}= 0.31,
$$
so the approximation based upon quasi-variances is quite accurate. 
In this case the $z$ statistic suggests that there is insufficient evidence to rule out the possibility that \emph{Biometrika} and \emph{Journal of the American Statistical Association} have the same ability to `export intellectual influence' within the $47$ Statistics journals in the list.

\begin{table}
\caption{\label{tab:BTrank} Journal ranking based on the Stigler model using data from Journal Citation Reports 2010 edition\null. Columns are the quasi-likelihood estimated Stigler-model  export scores (\texttt{SM}) with associated 
quasi standard errors (\texttt{QSE}), and estimated export scores
after grouping by lasso (\texttt{SM grouped}).}
\centering
\fbox{%
\begin{minipage}[t]{0.5\linewidth}
\begin{tabular}{rlccc}
Rank & Journal & SM & QSE & SM grouped \\ 
 \hline\hline
1 & JRSS-B & 2.09 & 0.11 & 1.87 \\ 
  2 & AoS & 1.38 & 0.07 & 1.17 \\ 
  3 & Bka & 1.29 & 0.08 & 1.11 \\ 
  4 & JASA & 1.26 & 0.06 & \dquote \\ 
  5 & Bcs & 0.85 & 0.07 & 0.65 \\ 
  6 & JRSS-A & 0.70 & 0.19 & 0.31 \\ 
  7 & Bern & 0.69 & 0.15 & \dquote \\ 
  8 & SJS & 0.66 & 0.12 & \dquote \\ 
  9 & Biost & 0.66 & 0.11 & \dquote \\ 
  10 & JCGS & 0.64 & 0.12 & \dquote \\ 
  11 & Tech & 0.53 & 0.15 & \dquote \\ 
  12 & AmS & 0.40 & 0.18 & 0.04 \\ 
  13 & JTSA & 0.37 & 0.20 & \dquote \\ 
  14 & ISR & 0.33 & 0.25 & \dquote \\ 
  15 & AISM & 0.32 & 0.16 & \dquote \\ 
  16 & CJS & 0.30 & 0.14 & \dquote \\ 
  17 & StSin & 0.29 & 0.09 & \dquote \\ 
  18 & StSci & 0.11 & 0.11 & -0.04 \\ 
  19 & LDA & 0.10 & 0.17 & \dquote \\ 
  20 & JRSS-C & 0.09 & 0.15 & \dquote \\ 
  21 & StMed & 0.06 & 0.07 & \dquote \\ 
  22 & ANZS & 0.06 & 0.21 & \dquote \\ 
  23 & StCmp & 0.04 & 0.15 & \dquote \\ 
  24 & StataJ & 0.02 & 0.33 & \dquote \\ 
\end{tabular}
\end{minipage}
\begin{minipage}[t]{0.5\linewidth}
\begin{tabular}{rlccc}
Rank & Journal & SM & QSE & SM grouped \\ 
  \hline\hline
  25 & SPL & -0.09 & 0.09 & -0.04 \\ 
  26 & StNee & -0.10 & 0.25 & \dquote \\ 
  27 & Envr & -0.11 & 0.18 & \dquote \\ 
  28 & JABES  & -0.16 & 0.23 & \dquote \\ 
  29 & Mtka & -0.18 & 0.17 & \dquote \\ 
  30 & StMod & -0.22 & 0.21 & \dquote \\ 
  31 & JSPI & -0.33 & 0.07 & -0.31 \\ 
  32 & SMMR & -0.35 & 0.16 & \dquote \\ 
  33 & BioJ & -0.40 & 0.12 & \dquote \\ 
  34 & JMA & -0.45 & 0.08 & -0.36 \\ 
  35 & EES & -0.48 & 0.25 &  \dquote \\ 
  36 & CSDA & -0.52 & 0.07 & \dquote \\ 
  37 & JNS & -0.53 & 0.15 & \dquote \\ 
  38 & CmpSt & -0.64 & 0.22 & \dquote \\ 
  39 & Stats    & -0.65 & 0.18 & \dquote \\ 
  40 & Test & -0.70 & 0.15 & \dquote \\ 
  41 & CSTM & -0.74 & 0.10 & \dquote \\ 
  42 & JSS & -0.80 & 0.19 & \dquote \\ 
  43 & JBS & -0.83 & 0.16 & \dquote \\ 
  44 & JSCS & -0.92 & 0.15 & \dquote \\ 
  45 & CSSC & -1.26 & 0.14 & -0.88 \\ 
  46 & StPap & -1.35 & 0.20 & \dquote \\ 
  47 & JAS & -1.41 & 0.15 & \dquote \\   
  &&&&\\
\end{tabular}
\end{minipage}
}
\end{table}

\subsection{Results}\label{sect:results}
We proceed now with interpretation of the ranking based on the Stigler model. It is reassuring that the four leading Statistics journals mentioned previously are ranked in the first four positions. \emph{Journal of the Royal Statistical Society} Series B is ranked first with a remarkably larger export score than the second-ranked journal, \emph{Annals of Statistics}: the approximate $z$ statistic for the significance  of the difference of their export scores is $5.44$\null.  The third position is occupied by \emph{Biometrika}, closely followed by \emph{Journal of the American Statistical Association}.

The fifth-ranked journal is \emph{Biometrics}, followed by \emph{Journal of the Royal Statistical Society} Series A, \emph{Bernoulli},   \emph{Scandinavian Journal of Statistics}, \emph{Biostatistics}, \emph{Journal of Graphical and Computational Statistics}, and \emph{Technometrics}.
 
\begin{figure}[!ht]
\centering
\makebox{\includegraphics[scale=.57]{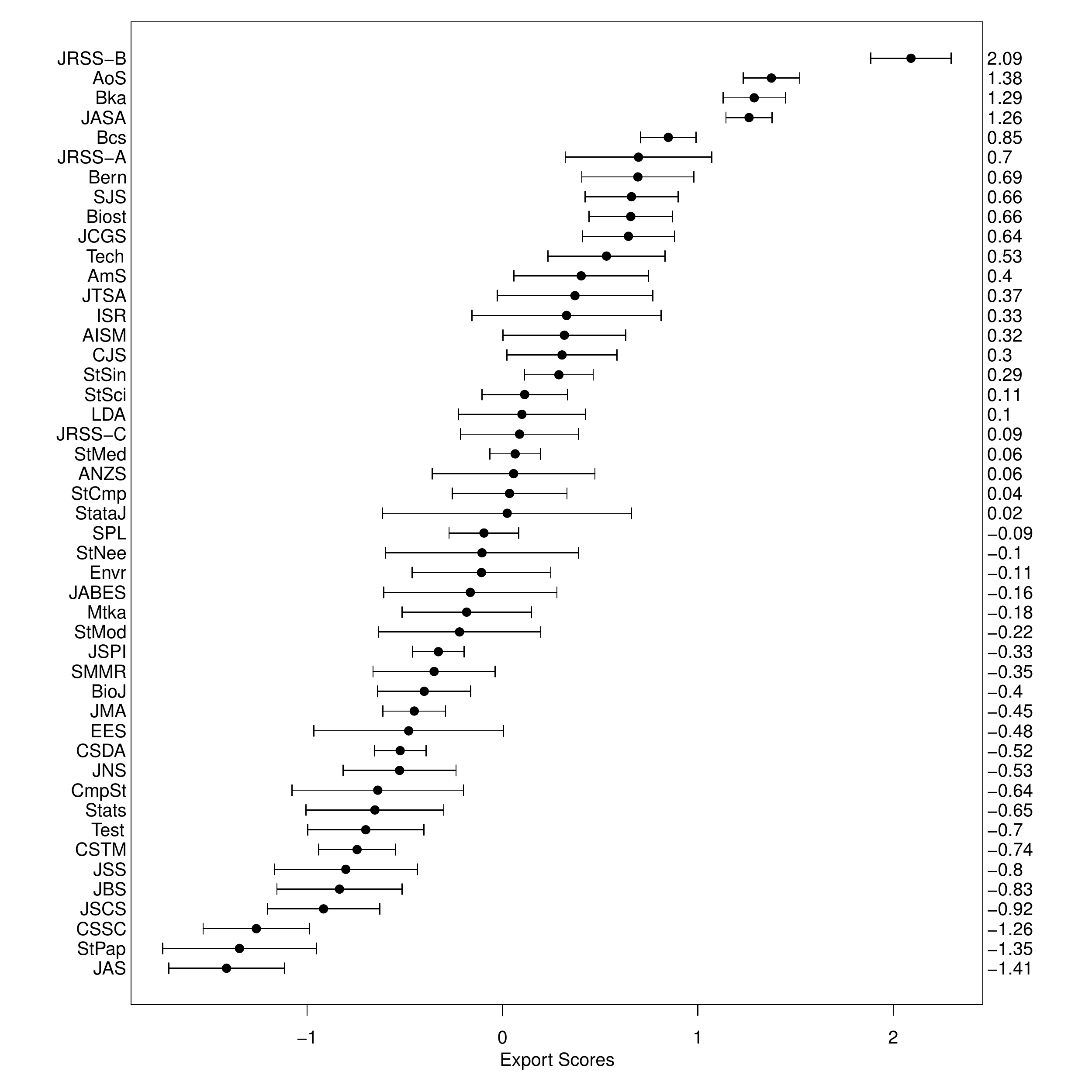}}
\caption{\label{fig:centipede}Centipede plot of 
estimated journal export scores and 
$95\%$ comparison intervals 
based on Journal Citation Reports 2010 edition\null. The error-bar limits are $\hat \mu_i \pm 1.96\, \text{qse}(\hat\mu_i)$, with the estimated export scores $\hat \mu_i$ marked by solid circles.}
\end{figure}

The `centipede' plot in Figure \ref{fig:centipede} visualizes the estimated export scores along with the $95\%$ comparison intervals with limits  $\hat \mu_i \pm 1.96\, \text{qse}(\hat\mu_i)$, where `qse' denotes the quasi standard error. The centipede plot highlights the outstanding position of \emph{Journal of the Royal Statistical Society} Series B, and indeed of the four top journals whose comparison intervals are well separated from those of the remaining journals. However, the most striking general feature is the substantial uncertainty in most of the estimated journal scores. Many of the small differences that appear among the estimated export scores are not statistically significant.  

\subsection{Ranking in groups with lasso}\label{sect:lasso}

Shrinkage estimation offers notable improvement over standard maximum likelihood estimation when the target is simultaneous estimation of a vector of mean parameters;
see, for example, \cite{morris_83}. It seems natural to consider shrinkage estimation also for the Stigler model.  \cite{masarotto_12} fit Bradley-Terry models with a lasso-type penalty \citep{tibshirani_96} which, in our application here, forces journals with close export scores to be estimated at the same level. The method, termed the ranking lasso, has the twofold advantages of shrinkage and enhanced interpretation, because it avoids over-interpretation of small differences between estimated journal export scores. 

For a given value of a bound parameter $s\geq 0$, the ranking lasso method fits the Stigler model by solving the quasi-likelihood equations (\ref{eq:quasi-likelihood})  with an $\text L_1$  penalty on all the pairwise differences of export scores; that is,
\begin{equation}\label{eqn:rlasso}
{\bf D}^\top {\bf V}^{-1} \left( \bm{c}- \bm{t} \bm{\pi}\right)=\bm{0}, \quad \text{subject to} \quad \sum_{i<j}^n w_{ij} | \mu_i-\mu_j|\leq s \quad \text{and} \quad \sum_{i=1}^n \mu_i=0,
\end{equation}
where the $w_{ij}$ are data-dependent weights discussed below.

Quasi-likelihood estimation is obtained for a sufficiently large value of the bound $s$. As $s$ decreases to zero, the $\text L_1$ penalty causes journal export scores that differ little to be estimated at the same value, thus producing a ranking in groups. The ranking lasso method can be interpreted as a generalized version of the fused lasso \citep{tibshirani_05}. 

Since quasi-likelihood estimates coincide with maximum likelihood estimates for the corresponding exponential dispersion model, ranking lasso solutions can be computed as penalized likelihood estimates. \cite{masarotto_12} obtain estimates  of the adaptive ranking lasso by using an augmented Lagrangian algorithm \citep{nocedal_06} for a sequence of bounds $s$ ranging from complete shrinkage ($s=0$) ---  \emph{i.e.},~all journals have the same estimated export score --- to the quasi-likelihood solution ($s=\infty$). 

Many authors (e.g.,~\citeauthor{fan_01}, \citeyear{fan_01}; \citeauthor{zou_06}, \citeyear{zou_06}) have observed that lasso-type penalties may be too severe, thus yielding inconsistent estimates of the non-zero effects. In the ranking lasso context, this means that if the weights $w_{ij}$ in (\ref{eqn:rlasso}) are all identical, then the pairwise differences $\mu_i-\mu_j$ whose `true' value is non-zero might not be consistently estimated. Among various possibilities, an effective way to overcome the drawback is to resort to the adaptive lasso method \citep{zou_06}, which imposes a heavier penalty on small effects. Accordingly, the adaptive ranking lasso employs weights equal to the reciprocal of a consistent estimate of $\mu_i-\mu_j$, such as
$ w_{ij}=| \hat \mu_i^{\text{(QLE)}} - \hat \mu_j^{\text{(QLE)}} |^{-1},$
with $\hat \mu_i^{\text{(QLE)}}$ being the quasi-likelihood estimate of the export score for journal $i$. 

Lasso tuning parameters are often determined by cross-validation. Unfortunately, the inter-journal `tournament' structure of the data does not allow the identification of internal replication, hence it is not clear how cross-validation can be applied to citation data. Alternatively, tuning parameters can be determined by minimization of suitable information criteria. The usual Akaike information criterion is not valid with quasi-likelihood estimation because the likelihood function is formally unspecified. A valid alternative is based on the Takeuchi information criterion (TIC; \citeauthor{takeuchi_76}, \citeyear{takeuchi_76}) which extends the Akaike information criterion when the likelihood function is misspecified. 
Let $\hat{\bm{\mu}}(s)=(\hat{\mu}_1(s), \ldots, \hat{\mu}_n(s))^{\top}$ denote the solution of (\ref{eqn:rlasso}) for a given value of the bound $s$. Then the optimal value for $s$ is chosen by minimization of
$$
\text{TIC}(s)=-2\,\hat\ell(s) + 2\, \text{trace}\left\{{\bf J}(s) {\bf I}(s)^{-1} \right\},
$$
where  $\hat\ell(s)=\ell\{\hat{\bm{\mu}}(s)\}$ is the misspecified log-likelihood of the Stigler model 
$$
\ell(\bm{\mu})=\sum_{i< j}^n c_{ij} (\mu_i-\mu_j) - t_{ij} \ln\{ 1+ \exp(\mu_i-\mu_j)\} 
$$
computed at $\hat{\bm{\mu}}(s)$, ${\bf J}(s)=\text{var}\{ \nabla \ell (\bm{\mu}) \}|_{\bm{\mu}={\hat{\bm{\mu}}(s)}}$ and ${\bf I}(s)=-\text{E}\{\nabla^2 \ell(\bm{\mu}) \}|_{\bm{\mu}={\hat{\bm{\mu}}(s)}}$. Under the assumed quasi-Stigler model, ${\bf J}(s)=\phi\, {\bf I}(s)$ and the TIC statistic reduces to 
$$
\text{TIC}(s)=-2\,\hat \ell(s) + 2\, \phi \, p,
$$
where $p$ is the number of distinct groups formed with bound $s$. The dispersion parameter $\phi$ can be estimated as in (\ref{eq:overdispersion}). The effect of overdispersion is inflation of the AIC model-dimension penalty. 
\begin{figure}[!h]
\centering
\makebox{
\includegraphics[scale=.55]{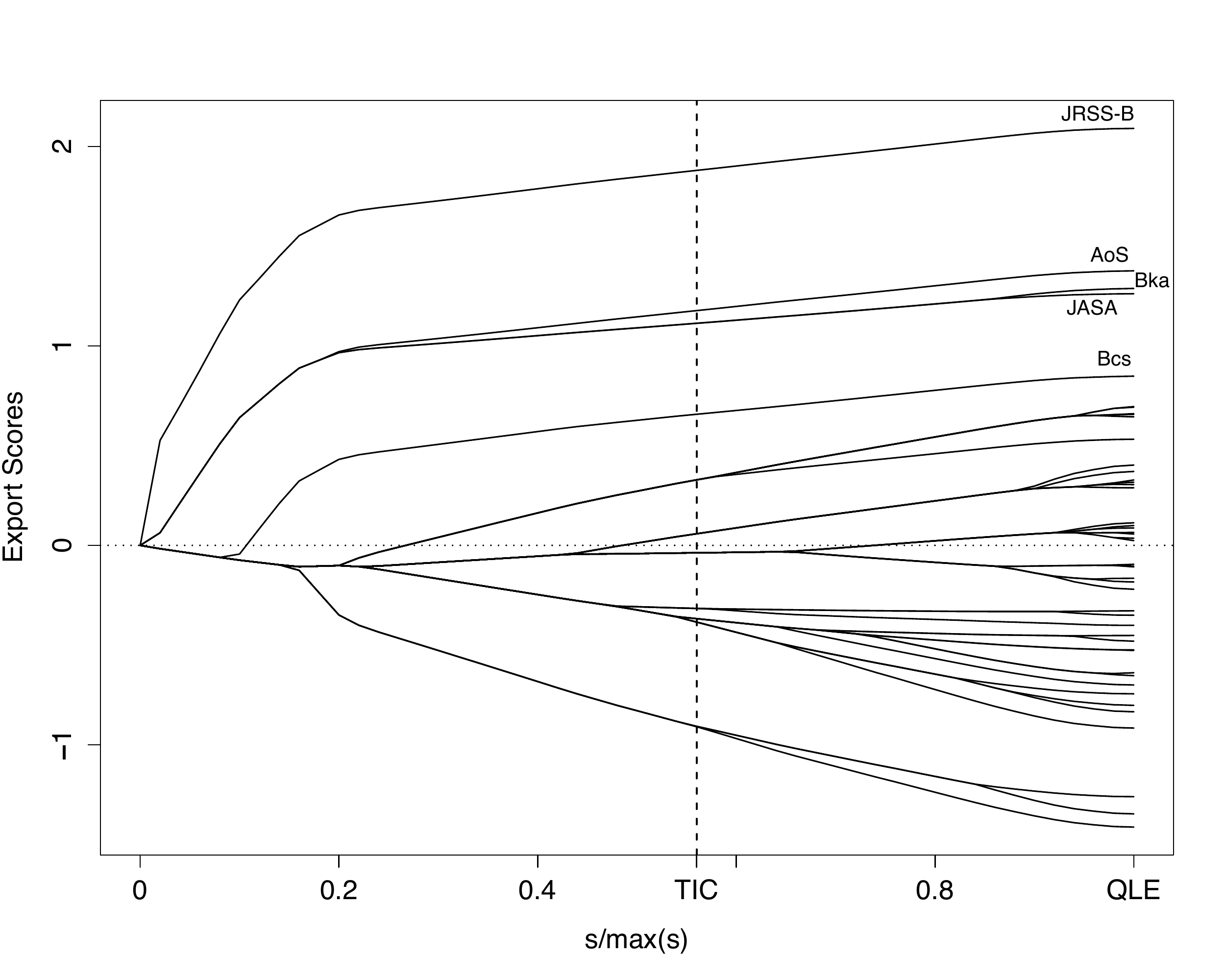}}
\caption{\label{fig:path}Path plot of adaptive ranking lasso analysis based on Journal Citation Reports 2010 edition.  {\tt QLE} indicates the quasi-likelihood estimate, {\tt TIC} the best ranking lasso solution according to the Takeuchi Information Criterion.}
\end{figure}

Figure \ref{fig:path} displays the path plot of the ranking lasso, while Table \ref{tab:BTrank} reports estimated export scores corresponding to the solution identified by TIC\null. See also Table \ref{tab:rankings} for a comparison with the Thomson Reuters published rankings. The path plot of Figure \ref{fig:path} visualizes how the estimates of the export scores vary as the degree of shrinkage decreases, \emph{i.e.}, as the bound $s$ increases. The plot  confirms the outstanding position of \emph{Journal of the Royal Statistical Society} Series B, the leader in the ranking at any level of shrinkage. Also \emph{Annals of Statistics} keeps the second position for about three-quarters of the path before joining the paths of \emph{Biometrika} and \emph{Journal of the American Statistical Association}\null. \emph{Biometrics} is solitary in fifth position for almost the whole of its path. The TIC statistic identifies a sparse solution with only 10 groups. According to TIC, the five top journals are followed by a group  of six further journals, namely \emph{Journal of the Royal Statistical Society} Series A, \emph{Bernoulli}, \emph{Scandinavian Journal of Statistics}, \emph{Biostatistics}, \emph{Journal of Computational and Graphical Statistics}, and  \emph{Technometrics}\null. However, the main conclusion from this ranking-lasso analysis is that many of the estimated journal export scores are not clearly distinguishable from one another.

\section{Comparison with results from the UK Research Assessment Exercise}\label{sect:RAE}

\subsection{Background}

In the United Kingdom, the quality of the research carried out in 
universities is assessed periodically by the 
government-supported funding councils, as a primary basis for 
future funding allocations.  At the time of writing, the most 
recent such assessment to be completed was the 2008 Research 
Assessment Exercise (RAE 2008), full details of which are online at 
\texttt{www.rae.ac.uk}.  The next such assessment to report, 
at the end of 2014, will be the similar `Research Excellence 
Framework' (REF).  Each unit of assessment is an academic 
`department', corresponding to a specified research discipline.  In
RAE 2008, `Statistics and Operational Research' was one of 67 such 
research disciplines; in contrast the 2014 REF has 
only 36 separate discipline 
areas identified for assessment, and research in Statistics will be
part of a new and much larger `Mathematical Sciences' 
unit of assessment.  The
results from RAE 2008 are therefore likely to provide 
the last opportunity
to make a directly Statistics-focused comparison with journal 
rankings.

It should be noted that the word `department' in RAE 2008 refers to
a discipline-specific group of researchers submitted for assessment 
by a university, or sometimes by two universities together: a 
`department' in RAE 2008 need not be an established academic unit 
within a university, and indeed many of the RAE 2008 Statistics 
and Operational Research `departments' were actually groups of 
researchers working in 
university departments of Mathematics or other disciplines.   

It is often argued that the substantial 
cost of assessing research outputs
through review by a panel of experts, as was done in RAE 2008,
might be reduced by employing suitable metrics based upon citation
data.  See, for example, \cite{jump_14}\null. 
Here we briefly explore this in a quite specific way, 
through data on journals rather than on the citations attracted by 
individual research papers submitted for assessment.  

The comparisons to be made here can also be viewed as exploring an 
aspect of `criterion validity' of the various journal-ranking 
methods: if highly ranked journals tend to contain high-quality
research, then there should be evidence through strong
correlations, even at the `department' level of aggregation, between
expert-panel assessments of research quality and 
journal-ranking scores.        

\subsection{Data and methods}

We examine only Sub-panel 22, `Statistics and Operational Research'
of RAE 2008\null.  The specific data used here are: 
\begin{itemize}
\item[(i)]
The detailed 
`RA2' (research outputs) submissions made by departments to RAE 2008\null. These list up to 4 research outputs per submitted researcher.
\item[(ii)]
The published RAE 2008 results on the assessed
quality of research outputs, 
namely the `Outputs sub-profile' for each department.
\end{itemize}

From the RA2 data, only research outputs 
categorized in RAE 2008 as
`Journal Article' are considered here.  For each such article, 
the journal's name is found in the `Publisher' field of the data.
A complication is that the name of any given journal can appear
in many different ways in the RA2 data, for example 
`Journal of the Royal Statistical Society B', 
`Journal of the Royal Statistical Society Series B: 
Statistical Methodology', \emph{etc.}; and the 
International Standard Serial Number (ISSN) codes as entered in 
the RA2 data are similarly unreliable.  Unambiguously resolving 
all of the many different 
representations of journal names proved
to be the most time-consuming part of the comparison exercise 
reported here.

The RAE 2008 `Outputs sub-profile' for each department gives the 
assessed percentage of research outputs at each of five quality levels,
these being `world leading' (shorthand code `4*'), 
`internationally excellent' (shorthand `3*'), then `2*', `1*' and 
`Unclassified'.  For example, the Outputs sub-profile for 
University of Oxford, 
the highest-rated Statistics and Operational Research submission in
RAE 2008, is
\begin{center}
\begin{tabular}{ccccc}
4* & 3* & 2* & 1* & U \\
37.0 & 49.5 & 11.4 & 2.1 & 0
\end{tabular}
\end{center}    
Our focus will be on the fractions at the 4* and 3* quality levels, 
since those are used as the basis for research funding.  
Specifically, in the 
comparisons made here the RAE `score' used will be
the percentage at 4* plus one-third of the percentage at 3*, 
computed from
each department's RAE 2008 Outputs sub-profile.  Thus, for example, 
Oxford's RAE 2008 score is calculated as $37.0 + 49.5/3 = 53.5$.  
This scoring formula is 
essentially the one used since 2010 to determine 
funding-council allocations; we
have considered also 
various other possibilities, such as simply the percentage at 4*, or
the percentage at 3* or higher, 
and found that the results below are not sensitive to this choice.

For each one of the journal-ranking methods listed in 
Table~\ref{tab:characteristics}, a bibliometrics-based comparator 
score per department is then constructed in a
natural way as follows.  Each RAE-submitted journal article
is scored individually, by for example the Impact Factor
of the journal in which it appeared; and those individual
article scores are then averaged across all of a 
department's RAE-submitted journal articles.  For the averaging,
we use the simple arithmetic mean of scores; an exception is that 
Stigler-model export scores are exponentiated prior to
averaging, so that they are positive-valued like the 
scores for the other methods considered.  Use of the median was
considered as an alternative to the mean; it was found to
produce very similar results, which accordingly will not be 
reported here.     

A complicating factor for the simple scoring scheme just described
is that journal scores were not readily available for all of
the journals named in the RAE submissions.  For the various `global'
ranking measures (cf.~Table~\ref{tab:characteristics}), scores 
were available for the 110 journals in the JCR 
\emph{Statistics and Probability} category, which covers 
approximately 70\% of the RAE-submitted journal articles 
to be scored.  For the Stigler model as used in this paper, though, 
only the subset of 47 Statistics journals listed in 
Table~\ref{tab:journals} are scored; and this subset accounts for 
just under 
half of the RAE-submitted journal articles.  In the following we 
have ignored all articles that appeared in un-scored journals,
and used the rest.  To enable a more direct comparison with the
use of Stigler-model scores, for each of the `global' indices
we computed also a restricted version of its mean score for each 
department, \emph{i.e.}, restricted to using scores for only
the 47 Statistics journals from Table~\ref{tab:journals}.

Of the 30 departments submitting work in
`Statistics and Operational Research' to RAE 2008, 4 
turned out to have substantially less than 50\% of their 
submitted journal articles in the JCR 
\emph{Statistics and Probability}
category of journals.  The data from those 4 departments, 
which were relatively small groups and whose RAE-submitted 
work was mainly in Operational Research, is omitted from the 
following analysis.  

The statistical methods used below to examine department-level 
relationships between the RAE scores and journal-based scores
are simply correlation coefficients and scatterplots.  Given 
the arbitrary nature of data-availability for this 
particular exercise, anything more sophisticated would seem 
inappropriate.

\subsection{Results}
\label{sect:RAE-results}

Table \ref{tab:RAE} shows, 
for bibliometrics-based mean 
scores based on each of the various journal-ranking 
measures discussed in this paper,
the computed correlation with departmental RAE score.
The main features of Table \ref{tab:RAE} are:
\begin{enumerate}
\item
The
Article Influence and Stigler Model scores correlate more
strongly with RAE results than do scores based on the other
journal-ranking measures.
\item
The various `global' measures show stronger correlation with
the RAE results when they are used only to score articles from
the 47 Statistics journals of Table \ref{tab:journals}, rather
than to score everything from the larger set of journals in 
the JCR \emph{Statistics and Probability} category.
\end{enumerate}
The first of these findings unsurprisingly 
gives clear support to the notion that
the use of bivariate citation counts, which take account of
the source of each citation and hence lead to measures of 
journal `prestige' rather than `popularity', is important if
a resultant ranking of journals should relate strongly to 
the perceived quality of published articles.  The second 
finding is more interesting: for good agreement with 
departmental RAE ratings, it can be substantially 
better to score only those
journals that are in a relatively homogeneous subset
than to use all of the 
scores that might be available for a larger set of 
journals.  In the present context, for example,
citation patterns for research in Probability are known to 
differ appreciably from those in Statistics, and `global' scoring of
journals across these disciplines would tend not to rate highly
even the very best work in Probability.

\begin{table}
\caption{\label{tab:RAE}
RAE 2008 score for research outputs in twenty-six UK `Statistics and Operational Research' departments: 
Pearson correlation with departmental mean
scores derived from the various 
journal-rating indices based on Journal Citation Reports 2010. 
}
\centering
\fbox{%
\begin{tabular}{l|cccccccc}
& \multicolumn{8}{c}{Journal Scoring Method} \\
Journals Scored & II & IF & IFno & IF5 &  AI & SM & SM grouped \\
\hline
&&&&&&&&\\
All of the JCR \emph{Statistics} & .34 & .47 & .49 & .50 & .73 & -- & -- \\
\emph{and Probability} category&&&&&&&&\\
&&&&&&&&\\
Only the 47 Statistics& .34 & .69 & .70 & .73 &   .79 & .81 & .82 \\
journals listed in Table \ref{tab:journals}&&&&&&&&\\
\end{tabular}
}
\end{table}

\begin{figure}
\centering
\makebox{\includegraphics[scale=.51]{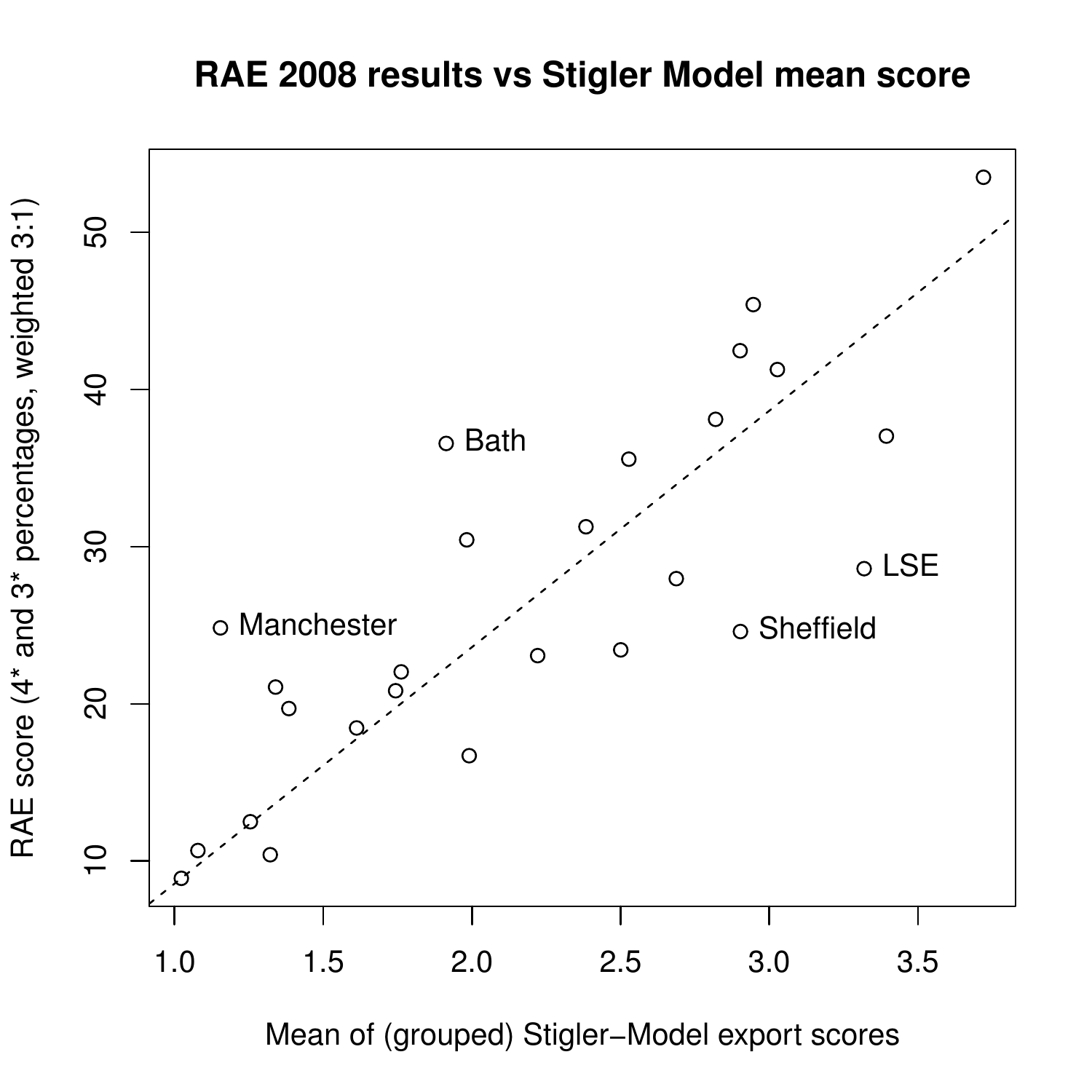}\includegraphics[scale=.51]{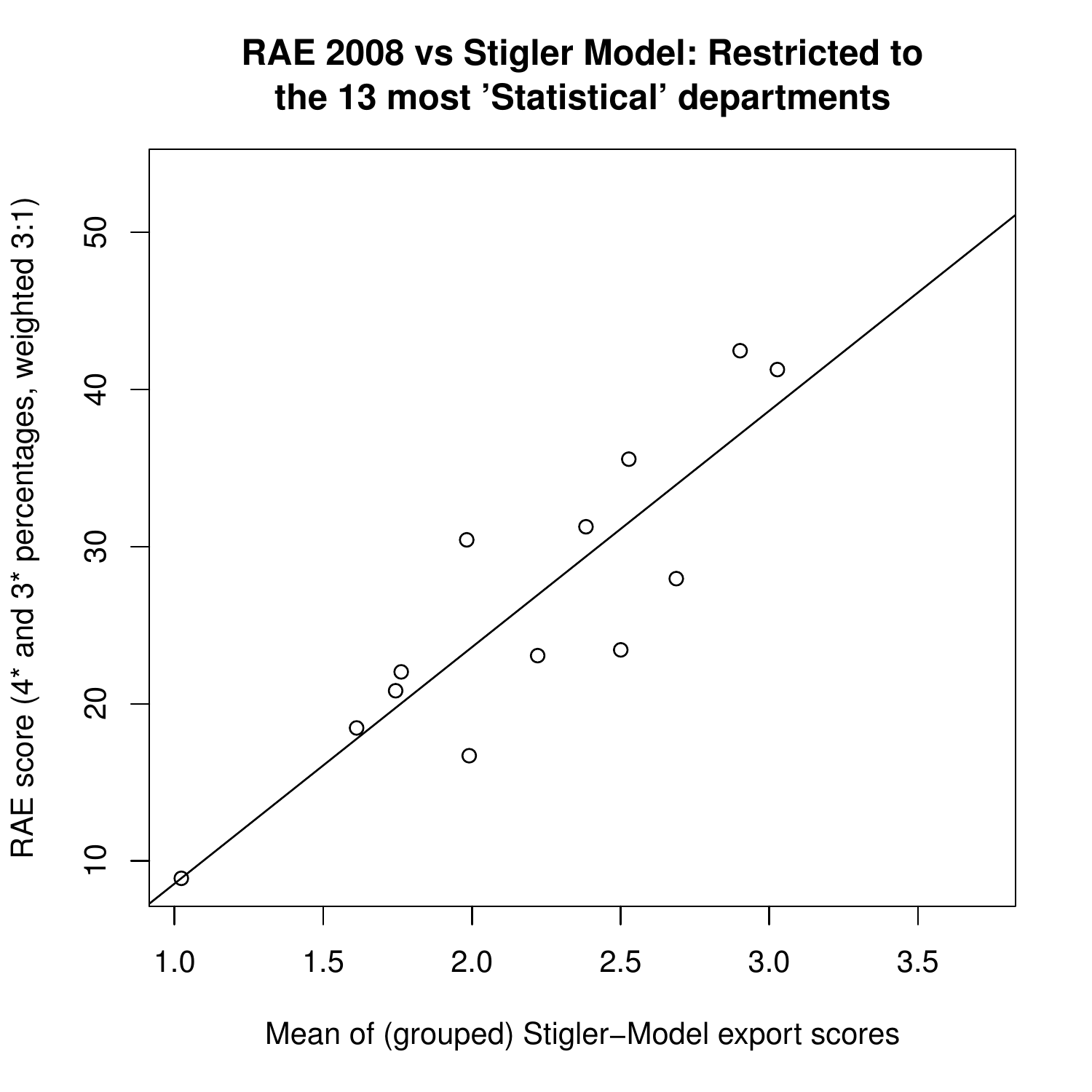}}
\caption{\label{fig:RAE}
Scatterplot (left panel) 
of RAE 2008 outcome 
(scores derived from the published RAE `Outputs' sub-profiles) 
against averaged 
Stigler Model journal export scores for RAE-submitted papers.  
The 26 plotted points
are the main `Statistics and Operational Research' groups at UK 
universities; four outliers from a straight-line fit are 
highlighted.  The right panel shows a subset of the same 
scatterplot, just the 13 research groups for which papers 
published in the 47 journals of Table \ref{tab:journals}
formed the majority of their RAE-submitted research outputs; the 
straight line shown in both panels is the least-squares fit to
these 13 points.}
\end{figure}

The strongest correlations found in Table \ref{tab:RAE} are those
based on journal export scores from the Stigler model, from columns
`SM' and `SM grouped' of Table~\ref{tab:BTrank}\null.   The 
departmental means of 
grouped export scores from the ranking-lasso method
correlate most strongly with RAE scores, a finding that supports
the notion that small estimated differences among 
journals are likely to be spurious.   
Figure \ref{fig:RAE} (left panel)
shows the relationship between RAE score and the mean of `SM grouped'
exponentiated journal export scores, for the 26 departments whose 
RAE-submitted journal articles were predominantly in the JCR
\emph{Statistics and Probability} category; the correlation 
as reported in Table \ref{tab:RAE} is 0.82\null.  The four largest 
outliers from a straight-line relationship are identified in the
plot, and it is notable that all of those four departments
are such that the ratio
\begin{equation}\label{eq:statfraction}
{
\textrm{Number of RAE outputs in the 47 Statistics journals of Table \ref{tab:journals}} \over
\textrm{Total number of RAE-submitted journal articles}
}
\end{equation}
is less than one-half.  
Thus the largest outliers are all departments
for which the majority of RAE-submitted journal articles are
not actually scored by our application of the Stigler model, and 
this 
seems entirely to be expected.  The right panel of 
Figure~\ref{fig:RAE} plots the same scores but now 
omitting all of the 13
departments whose ratio (\ref{eq:statfraction}) is less 
than one-half.  The result is, as expected, much closer to a 
straight-line relationship; the correlation in this restricted 
set of the most `Statistical' departments increases to 0.88\null.

Some brief remarks on interpretation of these findings 
appear in Section \ref{sec:metrics} below.  The data and
\texttt{R}-language code for this comparison are included
in this paper's Supplementary Web Materials.  
  
\section{Concluding remarks}\label{sect:conclusions}


\subsection{The role of statistical modelling in citation analysis}
In his Presidential Address at the {2011 Institute of Mathematical Statistics Annual Meeting} about controversial aspects of measuring research performance through bibliometrics, Professor Peter~Hall concluded that 
\begin{quote}
`As statisticians we should become more involved in these matters than we are. We are often the subject of the analyses discussed above, and almost alone we have the skills to respond to them, for example by developing new methodologies or by pointing out that existing approaches are challenged. To illustrate the fact that issues that are obvious to statisticians are often ignored in bibliometric analysis, I mention that many proponents of impact factors, and other aspects of citation analysis, have little concept of the problems caused by averaging very heavy tailed data. (Citation data are typically of this type.) We should definitely take a greater interest in this area'  \citep{hall_11}. 
\end{quote}
The model-based approach to journal ranking discussed in this paper is a contribution in the direction that Professor Hall recommended. Explicit statistical modelling of citation data has two important merits. First, transparency, since model assumptions need to be clearly stated and can be assessed through standard diagnostic tools. Secondly, the evaluation and reporting of uncertainty in statistical models can be based upon well established methods. 

\subsection{The importance of reporting uncertainty in journal rankings}

Many journals' websites report the latest journal Impact Factor and the journal's corresponding rank in its category. Very small differences in the reported Impact Factor often imply large differences in the corresponding rankings of Statistics journals. Statisticians should naturally be concerned about whether such differences are significant. Our analyses conclude that many of the apparent differences among estimated export scores are insignificant, and thus differences in journal ranks are often not reliable. The clear difficulty of discriminating between journals based on citation data is further evidence that the use of journal rankings for evaluation of individual researchers will often --- and perhaps always --- be inappropriate.

In view of the uncertainty in rankings, it makes sense to ask whether the use of
`grouped' ranks such as those that emerge from the lasso method of Section 
\ref{sect:lasso} should be universally advocated.  If the rankings or
associated scores are to be used for prediction purposes, then the usual 
arguments for shrinkage methods apply and such grouping, to help eliminate
apparent but spurious differences between journals, is likely to be beneficial;
predictions based on grouped ranks or scores are likely to be at least as
good as those made without the grouping, as indeed we found in 
Section~\ref{sect:RAE-results} in connection with RAE 2008 outcomes. For presentational purposes, though, the key requirement is at least some indication of the amount of uncertainty, and 
un-grouped estimates coupled with realistically wide intervals, as in the centipede plot of Figure~\ref{fig:centipede}, 
will often suffice.

\subsection{A `read papers' effect?}\label{sect:read_papers}

Read papers organised by the Research Section of the \emph{Royal Statistical Society} are a distinctive aspect of the \emph{Journal of the Royal Statistical Society} Series B\null. It is natural to ask whether there is a `read papers effect' which might explain the prominence of that journal under the metric used in this paper. During the study period 2001--2010, \emph{Journal of the Royal Statistical Society} Series B published in total 446 articles, 36 of which were read papers. Half of the read papers were published during the three years 2002--2004. The \emph{Journal of the Royal Statistical Society} Series B received in total 2,554 citations from papers published in 2010, with 1,029 of those citations coming from other Statistics journals in the list. Despite the fact that read papers were only 8.1\% of all published \emph{Journal of the Royal Statistical Society} Series B papers, they accounted for 25.4\% ($649/2554$) of all citations received by \emph{Journal of the Royal Statistical Society} Series B in 2010, and 23.1\% ($238/1029$) of the citations from the other Statistics journals in the list. 

Read papers are certainly an important aspect of the success of \emph{Journal of the Royal Statistical Society} Series B\null. However, not all read papers contribute strongly to the citations received by the journal. In fact, a closer look at citation counts reveals that the distribution of the citations received by read papers is very skew, not differently from what happens for `standard' papers. The most cited read paper published in 2001--2010 was \cite{spiegelhalter_02}, which alone received 11.9\% of all \emph{Journal of the Royal Statistical Society} Series B citations in 2010, and 7.4\% of those received from other Statistics journals in the list.  Some 75\% of the remaining read papers published in the study period each received less than  0.5\% of the 2010 \emph{Journal of the Royal Statistical Society} Series B citations. 

A precise quantification of the read-paper effect is difficult. Re-fitting the Stigler model dropping the citations received by read papers seems an unfair exercise.  Proper evaluation of the read-paper effect would require removal also of the citations received by other papers derived from read papers and published either in \emph{Journal of the Royal Statistical Society} Series B or elsewhere.

\subsection{Possible extensions}

\paragraph{Fractioned citations.} The analyses discussed in this paper are based on the total numbers $c_{ij}$ of citations exchanged by pairs of journals in a given period and available through the Journal Citation Reports. One potential drawback of this approach is that citations are all counted equally, irrespective of the number of references contained in the citing paper.  A number of recent papers in the bibliometric literature \citep[\emph{e.g.}, ][]{zitt_08, moed_10, leydesdorff_10, leydesdorff_11} suggest to re-compute the Impact Factor and other citation indices by using {fractional counting}, in which each citation is counted as $1/n$ with $n$ being the number of references in the citing paper. Fractional counting is a natural expedient to take account of varying lengths of reference lists in papers; for example, a typical review article contains many more references than does a short, technical research paper.  
The Stigler model extends easily to handle such fractional counting, for example through the quasi-symmetry formulation (\ref{eqn:quasisymmetry}); and the rest of the methodology described here would apply with straightforward modifications.

\paragraph{Evolution of export scores.}  This paper discusses a `static' Stigler model fitted to data extracted from a single JCR edition. A natural extension would be to study the evolution of citation exchange between pairs of journals over several years, through a dynamic version of the Stigler model.  A general form for such a model is 
$$
\text{log-odds}\left(\text{journal $i$ is cited by journal $j$ in year $t$}\right)=\mu_{i}(t)-\mu_{j}(t),
$$
where each journal's time-dependent export score $\mu_i(t)$ is assumed to be a separate, smooth function of $t$\null. 
Such a model would not only facilitate the systematic study of time-trends in the relative intellectual influence of journals, it would also `borrow strength' across years to help smooth out spurious variation, whether it be `random' variation arising from the allocation of citing papers to a specific year's JCR edition, or variation caused by transient, idiosyncratic patterns of citation. 
A variety of such dynamic extensions of the Bradley-Terry model have been developed in other contexts, especially the modelling of sports data; see, for example, \cite{fahrmeir_94}, \cite{glickman_99}, \cite{knorr-held_00} and \cite{cattelan_13}.

\subsection{Citation-based metrics and research assessment}
\label{sec:metrics}

From the strong correlations found in Section \ref{sect:RAE} 
between RAE~2008 outcomes and journal-ranking
scores, it is tempting to conclude that the expert-review element
of such a research assessment might reasonably be replaced, 
mainly or entirely, by automated scoring of journal articles
based on the journals in which they have appeared.   Certainly 
Figure \ref{fig:RAE} indicates that such scoring, when applied
to the main journals of Statistics, can perform 
quite well as a predictor of RAE outcomes for research groups
whose publications have appeared mostly in those journals.  

The following points should be noted, however:
\begin{enumerate}
\item
Even with correlation as high as 0.88, 
as in the right panel of Figure \ref{fig:RAE}, there can
be substantial differences between departments' positions
based on RAE outcomes and on journal scores. For example,
in the right panel of Figure \ref{fig:RAE} there are two departments whose
mean scores based on our application of the Stigler model 
are between 1.9 and 2.0 and thus essentially equal; but their
computed RAE scores, at 16.7 and 30.4, differ very substantially indeed.  
\item
High correlation was achieved by scoring only a relatively
homogeneous subset of all the journals in which the RAE-submitted
work appeared.  Scoring a wider set of journals, in order to
cover most or all of the journal articles appearing in the 
RAE 2008 `Statistics and Operational Research' submissions,
leads to much lower levels of agreement with
RAE results. 
\end{enumerate}
In relation to point (a) above it could of course be argued that,
in cases such as the two departments mentioned,
the RAE 2008 panel of experts got it wrong; 
or it could be that the difference seen between those two 
departments in the RAE results 
is largely attributable to the 40\% or so of journal articles
for each department that were not scored because they were outside
the list in Table \ref{tab:journals}\null.  Point (b), on
the other hand, seems more
clearly to be 
a severe limitation on the potential use of journal scores 
in place of expert review.  The use of cluster analysis
as in Section~\ref{sect:cluster}, in conjunction with expert
judgements about which journals are `core' to disciplines and sub-disciplines,
can help to establish relatively homogeneous subsets of journals
that might reasonably be ranked together; but comparison 
across the boundaries of such subsets is much more problematic.

The analysis described in this paper concerns journals. 
It says nothing directly about the possible use of citation data
on individual research outputs, as were made available 
to several of the review panels in the 2014 REF for example.  
For research in mathematics or statistics it seems clear that
such data on recent publications carry little information, mainly because of
long and widely-varying times taken for good research
to achieve `impact' through citations; indeed, 
the Mathematical Sciences 
sub-panel in REF 2014 chose not to use such data at all.  
Our analysis does, however, indicate that
any counting of citations to inform assessment of research quality
should at least take account of the source of each citation.

\section*{Acknowledgments}

The authors are grateful to Alan Agresti, Mike Titterington, the referees, the Series A Joint Editor and Associate Editor, and the Editor for Discussion Papers, for helpful comments on earlier versions of this work.  The kind permission of Thomson Reuters to distribute the JCR 2010 cross-citation counts is also gratefully acknowledged. 

This work was supported by the UK Engineering and Physical
Sciences Research Council through CRiSM grant EP/D002060/1, by
University of Padua grant CDPA131553, and by an IRIDE grant from
DAIS, Ca' Foscari University.

\end{document}